# Tensor Analysis and Fusion of Multimodal Brain Images

Esin Karahan, Pedro A. Rojas-Lopez, Maria L. Bringas-Vega, Pedro A. Valdes-Hernandez, Pedro A. Valdes-Sosa *

*Abstract*—Current high-throughput data acquisition technologies probe dynamical systems with different imaging modalities, generating massive data sets at different spatial and temporal resolutions posing challenging problems in multimodal data fusion. A case in point is the attempt to parse out the brain structures and networks that underpin human cognitive processes by analysis of different neuroimaging modalities (functional MRI, EEG, NIRS etc.). We emphasize that the multimodal, multi-scale nature of neuroimaging data is well reflected by a multi-way (tensor) structure where the underlying processes can be summarized by a relatively small number of components or "atoms". We introduce Markov-Penrose diagrams - an integration of Bayesian DAG and tensor network notation in order to analyze these models. These diagrams not only clarify matrix and tensor EEG and fMRI time/frequency analysis and inverse problems, but also help understand multimodal fusion via Multiway Partial Least Squares and Coupled Matrix-Tensor Factorization. We show here, for the first time, that Granger causal analysis of brain networks is a tensor regression problem, thus allowing the atomic decomposition of brain networks. Analysis of EEG and fMRI recordings shows the potential of the methods and suggests their use in other scientific domains.

## I. INTRODUCTION

WITHOUT doubt this is the era of "Big Data". Technological advances in data acquisition technology are spurring the generation of unprecedentedly massive datasets—thus posing a permanent challenge to data analysts. Recent international efforts are marshalling the use of a bewildering array of different technologies to acquire high-throughput multimodal information about real world systems. Examples of the systems and modalities probed are the internet [1], geophysical data [2], and the human genome [3].

The Human Brain is today perhaps the most challenging biological object under study and has been pushed recently to the forefront of public awareness [4]–[6]. This interest stems from the fact that identification of the brain structures involved in cognitive processes would not only yield essential understanding about the human condition, but would also provide leverage to deal with the staggering Global Burden of Disease for Brain Disorders [7]. The challenge of datasets stems from the fact that the brain is a highly nested set of interacting dynamical systems—from the subcellular level to the whole system. In turn, each level is actively being probed with an impressive arsenal of different measurement and imaging modalities. Some examples of system level measurements are high precision postmortem anatomy [8], diffusion weighted imaging (**DWI**) [9], [10], functional magnetic resonance imaging (**fMRI**) [11], electro encephalography (**EEG**) [12] and near infrared spectroscopy (**NIRS**) [13]. These are just some of the techniques that are rapidly populating publicly available databases (see for example those listed at www.incf.org). For an overview of the methods providing data see [14].

In this review brain information will serve us as an example of the complexities encountered in analyzing Big Data multimodal sets, as well as to illustrate possible strategies (**tensor models**) to cope with these issues. These approaches of course may be applied to other knowledge domains.

Typical issues that arise when analyzing neuroscience data are:
- Brain data is highly multidimensional and multimodal. Each imaging modality is always an indirect measurement of the underlying dynamical systems that are of interest. Thus we are required not only to solve multiple inverse problems (one for each measurement modality) but also to carry out multimodal fusion.
- Each modality is recorded at different spatial and temporal

"This work was supported in part by the TUBITAK-BIDEB 2214 International PhD and Postdoctoral Research Fellowship Programme for EK's visits to Cuban Neuroscience Center and the Key Laboratory for NeuroInformation of Ministry of Education, University of Electronic Science and Technology of China.

E. Karahan is with the Institute of Biomedical Engineering, Bogazici University, Kandilli Campus, 34684, Istanbul, Turkey (e-mail: esin.karahan@boun.edu.tr).

P. A. Rojas-Lopez and P. A. Valdes-Hernandez are with the Cuban Neuroscience Center, Ave 25 #15202 esq.158, Cubanacan, Playa, Cuba (e-mail: pedro.rojas@cneuro.edu.cu, multivac@cneuro.edu.cu).

M. L. Bringas-Vega is with the Key laboratory for NeuroInformation of Ministry of Education, Center for Information in BioMedicine, University of Electronic Science and Technology of China, 610054, Chengdu, China (email: maria@uestc.edu.cn, maluisabringas@yahoo.com)

P. A. Valdes-Sosa is with the Key laboratory for NeuroInformation of Ministry of Education, Center for Information in BioMedicine, University of Electronic Science and Technology of China, 610054, Chengdu, China (pedro@uestc.edu.cn) and the Cuban Neuroscience Center, Ave 25 #15202 esq.158, Cubanacan, Playa, Cuba (peter@cneuro.edu.cu) *corresponding author.

resolutions and reflects different physiological processes. This poses challenging problems for multimodal data fusion.
- To compound the complexity of brain data, the analysis is not only required to identify specific components and functions of the system but also to elucidate their interactions (i.e. to identify patterns of brain connectivity). This objective derives from consensus that neural computations are not the activity of relatively isolated neural masses but rather the coordinated activity of dynamically changing networks that involve huge amounts of neurons [15].

Though the methods we describe here for inverse problems and multimodal fusion are quite general, we will confine examples and detailed discussions to two types of brain imaging modalities, EEG and fMRI (BOLD). The choice of these is based on their common physiological basis and complementarity as we will immediately explain.

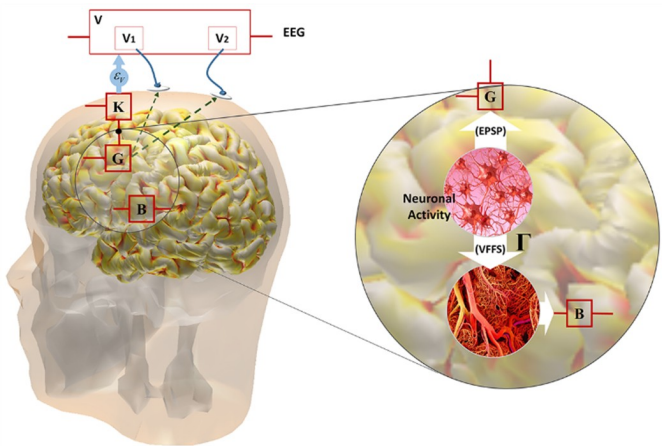

**Fig. 1.** Neural origin of the EEG and the BOLD signals. Neural activity in a cortical patch (amplified to the right of the figure) triggers two streams of events. On the one hand it induces an ensemble of post-synaptic potentials (EPSP) that creates primary current density **G** in gray matter. The second stream of events is the emission of $\Gamma$, a vasoactive feed forward signal (VFFS), that triggers a delayed and sluggish increase in blood flow resulting in an excess of oxy-hemoglobin which produces the BOLD signal **B** (fMRI). As shown in the left part of the figure, the cortical patch is embedded in a volume conductor (the head). As a consequence **G** is projected at electrodes ($V_1$ and $V_2$) placed at the scalp as determined by the lead field matrix **K** which encapsulates the effects of the volume conductor. The generation of **V** is further specified by use of the Markov-Penrose notation defined in this review (Fig. 4(d)). Modified from [23]. A detailed model concerning the generation of **B** can be found in Fig. 4(e).

As can be seen from Fig. 1 both types of measurements arise from neural activation which produces, on the one hand, primary current densities (**G**) that are reflected on the scalp as the EEG (**V**), and on the other hand as a vasoactive feed forward signal (VFFS, $\Gamma$) that produces a rush blood rich in oxy-hemoglobin that is measureable as a local **BOLD (blood oxygen level dependent)** signature (**B**) with magnetic resonance imaging, being the best known method of **fMRI**. The physical models describing the generation of **V** and **B** from **G** are known as the respective *forward problems*.

It is precisely due to the effect of these forward models that we must solve modality specific inverse in problems in order to uncover the underlying brain activation and connectivity. In the case of the EEG, the objective is to overcome the spatial smearing of the **G** that is due to volume conduction. For the EEG frequency ranges usually studied, the quasi-static approximation [16] guarantees that both forward and inverse problem are linear. EEG frequency components, or rhythms have been found to be important biomarkers of normal and abnormal brain states. A review of the basic facts about brain oscillations can be found in the comprehensive book [17].

In the case of the fMRI, the inverse problem compensates for the temporal blurring of the BOLD signal produced by the hemodynamic response function. Strictly speaking, the BOLD forward and inverse problems are nonlinear [18]. Nevertheless a useful first approximation is to linearize these equations (see [19] for a recent example).

Thus, in this review we shall be dealing with a dual set of linear forward and inverse problems. It is not a coincidence that much of the EEG and fMRI analysis literature read as exercises in the matrix analysis of ill posed linear inverse problems. This has been beneficial due to the large and rapidly expanding toolkit of matrix based methods analysis. As we shall show in Section IV, the indeterminacies of matrix decompositions have required additional assumptions in order to obtain well defined models tuned to the data. One such approach is Independent Component Analysis (ICA) which has become quite popular in both EEG and fMRI analysis [20] [21] [22].

Coming back to the generative models depicted in Fig. 1, it is clear that the EEG and fMRI have complementary strengths. EEG inverse solutions have low spatial resolution, while BOLD is weak in temporal granularity. Thus these modalities seem ideally matched for image fusion in order to create an brain mapping modality that simultaneously has high spatial and temporal resolution [23]–[25]. As with the modality specific inverse problems, a majority of the modeling strategies for this type of fusion is based on matrix methods [24], [26]–[29]. In particular, ICA methods for EEG/fMRI fusion are also very popular and have been recently reviewed in [30].

An alternative to purely matrix based EEG/fMRI inverse solutions and fusion methods is to take full advantage of the inherent structure of the multimodal brain data, which many times is actually recorded as a multidimensional array or *tensor*. It was early recognized [31] that tensors could be, under mild conditions, uniquely decomposed, a property applied with great advantage to the analysis of the EEG in the seminal work of [32], [33]. Following this lead, tensor based data analysis has been vigorously developed in the past few years [34]–[37]. The applications to neuroimaging data seem quite promising and have been extensively reviewed in [38]–[41]. Of course tensor methods have penetrated many fields of work. Several examples of their use in multimodal fusion outside of brain science can be found in [42].

Rather than replicating the material covered in the excellent reviews just mentioned above, our purpose here is to illustrate how tensor methods can enrich the statistical methodology underlying EEG/fMRI electrophysiological inverse solutions, brain connectivity, and image fusion. In order to do so we integrate and generalize work initiated in [23], [43]–[49]. To facilitate visual representation of models we introduce *Markov-Penrose diagrams* (**M-P diagram**), a combination of Penrose diagrams with notation from the theory of directed acyclic graphs (DAGs) that we use to represent Bayesian tensor models and clarify inferential steps.

We shall proceed as follows. Basic notation will be set down in Section II. This will allow us to first review matrix based EEG/fMRI analysis methods in Section III and to argue for the need of multi-linear approaches to the field. Section IV will present the Markov-Penrose diagrams as a convenient representation of tensors and operations. We then apply the ideas and notations of this section successively to EEG analysis (Section V); brain connectivity (Section VI); and EEG/fMRI fusion (Section VII). The descriptions will concentrate on model structure and results. We defer comments upon numerical algorithms to VIII, where we mention software packages for tensor based problems. Finally in section IX, we present some conclusions.

## II. NOTATION

### A. Tensor terminology

We selectively summarize here the notation used in this article, referring to the general texts for a more complete description. Tensors are a generalization of vectors and matrices to higher dimensions, being embodied in different programming languages as multidimensional arrays. The *order* of a tensor is the number of its dimensions. Thus:

- A scalar is a zeroth order tensor denoted by lower case symbols (e.g. $x$) except when denoting the number of levels in a given dimension, in which case, will be an upper case italic letter (e.g. $I$) without or with subscripts.
- A vector is a first order tensor denoted by lower case bold symbols (e.g. **x**).
- A matrix is a second order tensor denoted by upper case bold symbols (e.g. **X**).
- In its general form an N-Dimensional tensor (or an $N$ order tensor) is represented by calligraphic uppercase letters (e.g. $\mathcal{X} \in \mathbb{R}^{I_1 \times I_2 \times \cdots \times I_N}$).

In order to refer to subparts of tensors we shall use, when needed, MATLAB notation for multidimensional array indexing. Thus $\mathbf{X}(l,:)$ indicates the $l^{th}$ row of a matrix, $\mathcal{X}(:,:,l)$ is the frontal slice (or face) of a three dimensional tensor defined by fixing the third index.

$\mathbf{X}^T$ and $\mathbf{X}^H$ are the transpose and conjugate (Hermitian) transpose of the matrix $\mathbf{X}$, respectively. **I** is the identity matrix, *log* is the natural logarithm, and SVD is the singular value decomposition.

We shall also make use of the following operations (in Section VI) defined specifically for three dimensional tensors:

### B. Tensor t-operators

We now define a series of concepts and operations introduced by Kilmer and her group [50] that allows three dimensional tensors to be treated as though they were matrices. As we shall see in Section VI on brain connectivity three dimensions are adequate to define emitter, receiver and temporal signatures of nodes. In what follows $\breve{\mathcal{X}}$ will denote the Fourier transform of $\mathcal{X}$ along the third dimension.

- **The *MatVec* operation** matricizes a three dimensional tensor $\mathcal{X} \in \mathbb{R}^{I \times J \times K}$ as defined below:

$$\text{MatVec}(\mathcal{X}) = \begin{pmatrix} \mathcal{X}(:,:,1) \\ \mathcal{X}(:,:,2) \\ \vdots \\ \mathcal{X}(:,:,K) \end{pmatrix}$$

- **Tensor-product (t-product)** is a special type of product modified from [50]. Let $\mathcal{X}$ be a tridimensional tensor of size $I \times J \times K$ and $\mathcal{Y}$ of size $J \times L \times K$, then the t-product $\mathcal{X} *_t \mathcal{Y}$ is the $I \times L \times K$ tensor found by the carrying out the following operation for all *k* faces of the multiplicands $\breve{\mathcal{X}}(:,:,k) * \breve{\mathcal{Y}}(:,:,k)$. This multiplication is in effect a type of circular convolution along the third dimension.

- With this operation in place it is possible to define analogues for tensors of matrix concepts such as inverse, transpose, identity and diagonal tensors, orthogonal tensors and the singular value decomposition. We shall refer to these operators as with their usual name preceded by a t: e.g. t-inverse.

- The **t-SVD** of $\mathcal{X} \in \mathbb{R}^{I \times J \times K}$ is: $\mathcal{X} = \mathcal{U} *_t \mathcal{D} *_t \mathcal{V}^T$, where $\mathcal{U} \in \mathbb{R}^{I \times I \times K}$, $\mathcal{V} \in \mathbb{R}^{J \times J \times K}$ are t-orthogonal tensors and $\mathcal{D} \in \mathbb{R}^{I \times J \times K}$ is a tensor with diagonal faces. As with the usual SVD it provides an optimal approximation of a tensor in the Frobenius norm of the difference. (see Theorem 4.3 in [51]).

- The **tensor nuclear norm** (TNN) is defined from the t-SVD as: $\|\mathcal{X}\|_\circledast = \sum_{i=1}^{\min(I,J)} \sum_{k=1}^{K} \breve{\mathcal{D}}(i,j,k)$

*t-product* operators can be efficiently computed by taking means of the Fourier transforms of tensors along the third mode. [50].

**The *Tplz* operation:**

A three dimensional tensor $\mathcal{X} \in \mathbb{R}^{I \times J \times K}$ can be converted into a block toeplitz matrix of dimension $I \cdot K \times J \cdot K$ by rearrangement of its faces as follows:

$$\text{tplz}(\mathcal{X}) = \begin{pmatrix} \mathcal{X}(:,:,1) & \mathcal{X}(:,:,2)^H & \cdots & \mathcal{X}(:,:,K)^H \\ \mathcal{X}(:,:,2) & \mathcal{X}(:,:,1) & \cdots & \mathcal{X}(:,:,K-1)^H \\ \vdots & \vdots & \ddots & \vdots \\ \mathcal{X}(:,:,K) & \mathcal{X}(:,:,K-1) & \ddots & \mathcal{X}(:,:,1) \end{pmatrix}$$

*TABLE I. Symbols used in the article*

| Dimension | Definition |
|---|---|
| $I_E$ / $I_{E\delta}$ | Number of EEG scalp electrodes * |
| $I_{Cx}$ / $I_{Cx\delta}$ | Number of sources (EEG or BOLD) on a cortical surface grid |
| $I_{Wm}$ | Number of voxels in white matter tracts |
| $I_T$ / $I_{T\delta}$ | Number of time points |
| $I_F$ / $I_{F\delta}$ | Number of frequency points |
| $I_W$ | Number of subjects |
| $I_{lag}$ | Number of past time points (time lags) in Autoregressive models |
| $N$ | Number of dimensions of a tensor |
| $R$ | Number of atoms (components) in tensor decompositions |

*Subscript $\delta$ is used for subsampling of that dimension

*TABLE II Matrices and tensors*

| Symbol | Definition | Dimension |
|---|---|---|
| **Constants** | | |
| **K** | EEG Lead field | $I_E \times I_G$ |
| **G** | Primary current density matrix (Generators) of EEG | $I_{Cx} \times I_T$ |
| **H** | Hemodynamic response matrix | $I_T \times I_{T\delta}$ |
| **Γ** | Vasoactive feed-forward signal matrix (VFFS) | $I_{Cx} \times I_T$ |
| **L** | Laplacian matrix | $I_E \times I_E$ or $I_{Cx} \times I_{Cx}$ |
| **Data tensors** | | |
| **V** | EEG space/time data matrix | $I_E \times I_T$ |
| **B** | fMRI matrix | $I_{Cx} \times I_{T\delta}$ |
| **FA** | DTI-FA data matrix | $I_{Wm} \times I_S$ |
| **B**$_t$ | Data matrix of Granger Causality (GC) | $I_{Cx} \times I_T$ |
| $\mathcal{B}$ | Time lagged data tensor of GC | $I_{lag} \times I_{Cx} \times I_T$ |
| **Estimated tensors** | | |
| $\mathcal{S}_T$ | Space/time/frequency tensor for spectral estimate of EEG | $I_E \times I_{T\delta} \times I_{\delta F}$ |
| $\mathcal{S}_S$ | Space/subject/frequency tensor for spectral estimate of EEG | $I_E \times I_S \times I_{\delta F}$ |
| $\mathcal{A}$ | GC Connectivity tensor | $I_{Cx} \times I_{Cx} \times I_{lag}$ |
| **Signatures of tensor decompositions** | | |
| **M**$_V$ | Spatial signature of EEG atoms over electrodes | $I_E \times R$ |
| **M**$_G$ | Spatial signature of EEG over the cortical grid | $I_{Cx} \times R$ |
| **M**$_B$ | Spatial signature of fMRI atoms over the cortical grid | $I_{Cx} \times R$ |
| **M**$_{FA}$ | Spatial signatures of FA atoms over the white matter grid | $I_{Wm} \times R$ |
| **T**$_V$ | Temporal signatures of EEG for atoms | $I_{T\delta} \times R$ |
| **T**$_B$ | Temporal signatures of fMRI atoms for sampled time points | $I_{T\delta} \times R$ |
| **F**$_V$ | Spectral signatures of EEG atoms | $I_{F\delta} \times R$ |
| **W**$_B$ | Subject signatures of fMRI atoms | $I_S \times R$ |
| **W**$_{FA}$ | Subject signatures for FA atoms | $I_S \times R$ |
| **M**$_r$ | Spatial signatures for receiver voxels in GC atoms | $I_{Cx} \times R$ |
| **M**$_t$ | Spatial signatures for sender voxels in GC atoms | $I_{Cx} \times R$ |
| **T** | Temporal signatures for GC atoms | $I_{lag} \times R$ |

## III. MATRIX BASED EEG/FMRI ANALYSIS

We now formalize the EEG forward and inverse problems outlined in the Introduction and presented in Fig 1. Both the EEG and fMRI are vector valued time series, collected into matrices (i.e. space × time).

Let us represent the recorded EEG by the matrix $\mathbf{V} \in \mathbb{R}^{I_E \times I_T}$, where $I_E$ denotes the number of electrodes in the recording sensor array placed on the scalp, and $I_T$ denotes the total number of observations obtained during the recording epoch. The sampling rate of the EEG is typically in the range of around 1 KHz.

BOLD measurements are available for all brain voxels in an image but we will assume that standard preprocessing techniques have allowed us to project this activity to a more restricted space. This can be either a standard 2-D grid distributed on the cerebral cortex, or a larger set defined for a 3-D grid that spans cerebral cortex and thalamus. The distinction between the two cases will be clear from the context but in both cases with a total number of voxels $I_{Cx}$. This activity is sampled at much slower rates than the EEG (typically in the order of 1 Hz) which results in a total number of observations $I_{T\delta} \ll I_T$. We therefore denote the recorded fMRI by the matrix $\mathbf{B} \in \mathbb{R}^{I_{Cx} \times I_{T\delta}}$.

We also assume for EEG and fMRI recorded concurrently, that the time samples for the fMRI are obtained at integer multiples of those for the EEG.

### A. The Matrix EEG inverse problem

The discretized version of the forward problem for the EEG is:

$$\mathbf{V} = \mathbf{KG} + \varepsilon_V \tag{1}$$

where $\mathbf{G} \in \mathbb{R}^{I_{Cx} \times I_T}$ denotes the primary current density defined over the same cortical grid as the fMRI and sampled at the same time points as the EEG. $\mathbf{K} \in \mathbb{R}^{I_E \times I_{Cx}}$ is the lead field matrix which summarizes volume conduction effects in the head[1]. We shall henceforth assume that that the sensor error $\varepsilon_V \in \mathbb{R}^{I_E \times I_T}$ is a matrix with entries that are identically distributed zero mean Gaussian variates. This assumption allows us to measure the fit of models to the data by means of the Frobenius norm $\|\bullet\|_2^2$.[2]

Estimation of $\mathbf{G}$, also known as Electrophysiological Source Imaging (ESI), is a well-known ill posed problem. Therefore a solution by naive minimization of the functional $\|\mathbf{V} - \mathbf{KG}\|_2^2$ is not possible and, in fact, does not have a unique solution. Uniqueness may be obtained by adding prior anatomical and physiological information to the problem formulation.

According to this approach, estimation of $\mathbf{G}$ involves finding the argument $\hat{\mathbf{G}}$ that minimizes the following augmented functional:

$$\hat{\mathbf{G}} = \arg\min_{\mathbf{G}} \|\mathbf{V} - \mathbf{KG}\|_2^2 + \pi(\mathbf{G}) \quad (2)$$

The penalization term $\pi(\mathbf{G})$ applied in ESI is generally a combination of different matrix norms. Let us consider some examples.

- One of the best known example of source imaging is LORETA [52], in which the penalization takes the form $\pi(\mathbf{G}) = \|\mathbf{LG}\|_2^2$, that encourages estimation of spatially smooth sources on the cortex.
- Another variant is VARETA [53] which uses the penalty term $\pi(\mathbf{G}) = \|\mathbf{LG}\|_1$ with an L1 norm to impose both spatial smoothness and sparseness.
- More recently in [49], the following penalty term was proposed $\pi(\mathbf{G}) = \lambda_1 \|\mathbf{G}\|_1 + \lambda_2 \|\mathbf{LG}\|_2^2$. This penalty achieves an optimal balance between spatial sparsity and smoothness of cortical sources by data driven hyper-parameters $\lambda_1$ and $\lambda_2$.

A comparison of these and other types of source imaging may be found in [48]. In that same paper STONNICA (**S**patio-**T**emporal **O**rthogonal **N**on **N**egative **I**ndependent **C**omponent **A**nalysis) was proposed as a solution for the EEG inverse problem. This model not only illustrates more complex penalty terms, but also shows how additional constraints might be useful to find interpretable sources. A tensor extension is described in Section V.

STONNICA is different in two ways from usual applications of ICA to EEG source localization [20].
1. It is based on a variant of ICA for which components are forced to be Orthogonal and Non-Negative (ONN-ICA) [54]
2. STONNICA identifies the components in source space directly (ICA tomography). By contrast, other types of ICA source/localization first carries out ICA in sensor space and then localizes the extracted components (tomography of ICA).

The parameter estimates for this model are obtained as:

$$(\hat{\mathbf{M}}_V, \hat{\mathbf{T}}_V) = \arg\min_{\mathbf{M}_V, \mathbf{T}_V, \mathbf{F}_V} \left\{ \begin{array}{l} \frac{1}{2} \|\mathbf{V} - \mathbf{K}\mathbf{M}_V \mathbf{T}_V\|_2^2 \\ + \lambda_1 \|\mathbf{M}_V\|_1 + \lambda_2 \|\mathbf{L}\mathbf{M}_V\|_2^2 \end{array} \right\} \quad (3)$$
$$\text{s.t.} \mathbf{M}_V^T \mathbf{M}_V = \mathbf{I}, \mathbf{M}_V \geq 0$$

where $\mathbf{M}_V \in \mathbb{R}^{I_{Cx} \times R}$ is the spatial signature or cortical distribution of the ONN-ICA components. The ONN condition is equivalent to specifying spatially non-overlapping EEG sources. This requirement, in addition to the smooth Lasso type [44], [46], [49], results in the identification of sparse isolated clustered components that were used to identify distinct cognitive processes involved in face processing.

### B. The Matrix fMRI Inverse problem

As mentioned before, while the generation of the BOLD signal from the VFFS is best described by a set of nonlinear differential equations [18], strictly speaking the forward and inverse fMRI problem should be solved using neural mass models based on nonlinear random differential equations as reviewed in detail in [23]. But this would detract from our objective of simplicity in illustrating matrix and tensor techniques.

We shall therefore resort to a useful linear generative model for the BOLD signal which is the convolution of the vasoactive feed forward signal, $\gamma(g,t)$ at point $g$ of the cortical grid with the hemodynamic response $h(t)$:

$$b(g,t) = \int h(t-\tau) \gamma(g,t) d\tau + \varepsilon_b(g,t) \quad (4)$$

If this continuous time model is discretized over time at the $I_{T\delta}$ sampling times of the fMRI and the convolution operation stated as a matrix product, the resulting fMRI forward model is:

$$\mathbf{B} = \mathbf{\Gamma H} + \varepsilon_B \quad (5)$$

where $\mathbf{\Gamma} \in \mathbb{R}^{I_{Cx} \times I_T}$ is the vasoactive feed-forward signal matrix and $\mathbf{H} \in \mathbb{R}^{I_T \times I_{T\delta}}$ is the hemodynamic response matrix. Note that $\mathbf{H}$ is obtained by subsampling rows from the general square symmetric Toeplitz matrix $\{h(t_i - t_j)\}_{1 \leq i, j \leq I_T}$ defined at the finer time resolution.

Using similar procedures as those outlined in the discussion of ESI, the deconvolution of the fMRI may also be stated as an inverse problem:

$$\hat{\mathbf{\Gamma}} = \arg\min_{\mathbf{\Gamma}} \|\mathbf{B} - \mathbf{\Gamma H}\|_2^2 + \pi(\mathbf{\Gamma}) \quad (6)$$

---

[1] Note that this is a linear operator under the quasi-static approximation for Maxwell's equations, [140].

[2] Correlated or non-Gaussian error terms can easily be dealt with, but would complicate model expressions unnecessarily.

Glover [55] was the first to propose this type of deconvolution using a Wiener filter, that is with $\pi(\mathbf{\Gamma}) = \|\mathbf{\Gamma}\|_2^2$. Later [23] proposed the use of the penalty $\pi(\mathbf{\Gamma}) = \|\mathbf{\Gamma}\mathbf{L}\|_2^2$ (a "LORETA style" inverse problem) in the context of EEG/fMRI fusion, a topic which we now turn to.

### C. Matrix based EEG/fMRI fusion

From an examination of (2) and (6) it is clear that in order to carry out EEG/fMRI fusion a link must be established between $\mathbf{G}$ and $\mathbf{\Gamma}$. A first attempt was carried out by [56] in which both quantities where assumed to be proportional to each other. Under these conditions a form of matrix EEG/fMRI fusion procedure was developed that was formulated as:

$$\hat{\mathbf{G}} = \arg\min_{\mathbf{G}} \|\mathbf{V} - \mathbf{KG}\|_2^2 + \alpha \|\mathbf{B} - \mathbf{GH}\|_2^2 + \pi(\mathbf{G}) \quad (7)$$

Using LORETA type penalties this method was capable of accurately localizing separate components of the Somatosensory Evoked Magnetic Response. Also note that this was the first formulation of a symmetrical type of EEG/fMRI fusion, since neither modality has a priori priority over the other—their relative weight is determined by the data chosen constant $\alpha$. Symmetrical data fusion approaches use complementary information of both modalities to estimate the common source of neural activity [56], [57] whereas in asymmetrical fusion methods one modality is given priority to guide the other one. A review can be found in [58].

Another type of matrix based fusion is extracting the features of two modalities such as contrast maps for fMRI and averaged time course of EEG and apply fusion methods on second statistics to reveal the between subject variance. For example, in joint ICA, data features matrices are concatenated and decomposed by assuming that the datasets share the same mixing matrix/modulation profiles but different source components. In [59], the temporal data feature matrix of EEG ($\mathbf{V}$) constituted by the average ERP time courses over subjects and contrast maps of fMRI over subjects ($\mathbf{B}$) are jointly decomposed to find the temporal ($\mathbf{T_V}$) and spatial components ($\mathbf{M_B}$) as stated below:

$$[\mathbf{V}, \mathbf{B}] = \mathbf{A}[\mathbf{T_V}, \mathbf{M_B}]$$
$$\text{s.t. } \pi(\mathbf{T_V}) = \prod_{r=1}^{R} \pi(\mathbf{T_V}(r,:)), \; \pi(\mathbf{M_B}) = \prod_{r=1}^{R} \pi(\mathbf{M_B}(r,:))$$

where $\mathbf{A}$ is the common modulation profiles of subjects and $R$ is the number of components of ICA.

Other, similar examples of multimodal fusion can be found in [25].

### D. Beyond Matrix Based Methods

As we have just seen, matrix based methods have been very useful for EEG/fMRI inverse solutions and their fusion. Nevertheless many EEG and fMRI datasets and model constructs are not best represented as matrices. Consider a set of EEG spectra recorded on a common set of sensors, for a number of subjects, for different experimental conditions. This is actually the data that can be arranged as a multidimensional array (derivation, frequency, subject, condition).

Instead of reshaping the data into matrices for use with more standard methods, tensor decomposition leverages the natural format of the data in order to discover hidden structures (see [60]). It is therefore not surprising to see the waxing of tensor methods in the analysis of brain data [61] and [39].

## IV. TENSOR DIAGRAMS AND OPERATIONS

### A. Penrose Diagrams

Tensor objects and operations may be represented by Penrose diagrams (**P diagram**) that were inspired by Penrose's original work in [62] in theoretical physics and later adapted, with modifications, for tensor networks. Examples of P diagrams can be found in [41], [63], [64]. Expressions in usual mathematical notation for vectors, matrices and tensors as well as their corresponding Penrose diagrams are shown in Fig 2(a,b,c). In this type of diagram, tensors are denoted as nodes and each dimension is depicted as a line (*link*) leaving the node. The model order of a tensor can be read off from its Penrose diagram as the number of dangling lines. Note that any tensor may have any number of additional "singleton" or virtual one-element dimensions (e.g. $\mathcal{X} \in \mathbb{R}^{I_1 \times I_2 \times \cdots \times I_N}$ is the same tensor as $\mathcal{X} \in \mathbb{R}^{I_1 \times I_2 \times \cdots \times I_N \times 1 \times \cdots \times 1}$). Tensors which have random values as elements are shown as circular nodes. Other nodes with special properties are depicted as in Fig 2. Those with constant/deterministic elements are shown as rectangular nodes (Fig 2(d)). If all of the elements of a tensor are nonnegative this property (or for that matter any other relation) is depicted within the node outline (Fig 2(e)). A tensor with orthogonality imposed on a given dimension will be shown with a rectangular box on that dimension (Fig 2(f)).

Any junction between lines of two or more tensors denotes an operation between them that may result in a lower order tensor.

We now list those tensor operations used in this review, though of course there are many more. These are:

- **Tensor contraction** [See Fig. 3(a,b)] which is a generalization of matrix multiplication for tensors. Consider a tensor $\mathcal{X}$ of size $I_1 \times \cdots \times I_N \times J_1 \times \cdots \times J_M$ and $\mathcal{Y}$ of size $J_1 \times \cdots \times J_M \times K_1 \times \cdots \times K_P$. Contraction over common dimensions $J_1 \times \cdots \times J_M$ will give:

$$(\mathcal{X} \bullet_{\{j_1,\cdots,j_M\}} \mathcal{Y})(i_1,\cdots,i_N,k_1,\cdots,k_P)$$
$$= \sum_{j_1\cdots j_M=1}^{J_1\cdots J_M} \mathcal{X}(i_1,\cdots,i_N,j_1,\cdots,j_M)\mathcal{Y}(j_1,\cdots,j_M,k_1,\cdots,k_P)$$

- **Tensor concatenation** is the merging of two tensors whose dimensions should be same except for the

concatenation index. Consider $\mathcal{X}$ of size $I_1 \times \cdots I_{n-1} \times J \times I_{n+1} \cdots \times I_N$ and $\mathcal{Y}$ of size $I_1 \times \cdots I_{n-1} \times K \times I_{n+1} \cdots \times I_N$, concatenation of these tensors on the $J^{th}$ and $K^{th}$ dimensions will give the tensor $\mathcal{Z} = [\mathcal{X}, \mathcal{Y}]$ of size $I_1 \times \cdots I_{n-1} \times J+K \times I_{n+1} \cdots \times I_N$. Note that tensor concatenation preserves tensor order. See Fig. 3(c,d) for an example.

- **The Kruskal operator** [65], [66] is an important operation for constructing a tensor from matrices with the same number of columns $R$. For a three dimensional tensor the product is defined as $\mathcal{X} = [\![\mathbf{A}, \mathbf{B}, \mathbf{C}]\!] = \mathbf{A} \bullet_{\{R\}} \mathbf{B} \bullet_{\{R\}} \mathbf{C}$. The famous PARAFAC model is defined in terms of this operator, $R$ being the number of **atoms** and each of the matrices characterizing a certain **signature**. (All of the factor matrices are normalized to unit norm except for the first matrix.) Fig. 5 shows an example of an M-P diagram for a PARAFAC model.

For clarity, we sometimes emphasize a particular tensor, the result of several operations, by surrounding it with triangular or rectangular shapes as shown in Fig. 3(e). Note that the dimensionality of the resultant tensor will be clear from the lines leaving the shape.

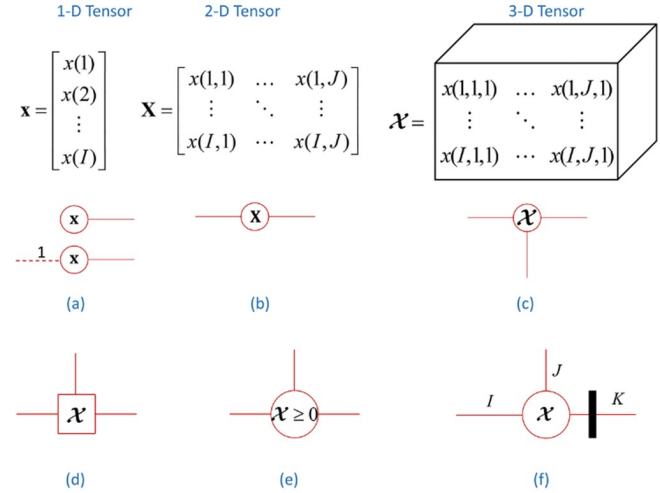

**Fig. 2.** Penrose Diagrams (a,b,c,d) Graphical depictions of tensors of different orders (i.e. number of dimensions) as circles, when referring to unobserved variables, and as squares, when denoting constant tensors. The symbol used to denote the tensor is contained in the circle or square. Note that the number of dimensions of a tensor is shown as the lines that leave it. When necessary, the number of elements of a dimension will be indicated beside the link that denotes it. Singleton dimensions (dimensions with only one level) are shown as dashed lines with the number 1. (e) Nonnegativity and other relations for tensor elements indicated with the usual mathematical notation in the circle or square. (f) Orthogonality constraints are shown as squared bars on the orthogonal dimensions.

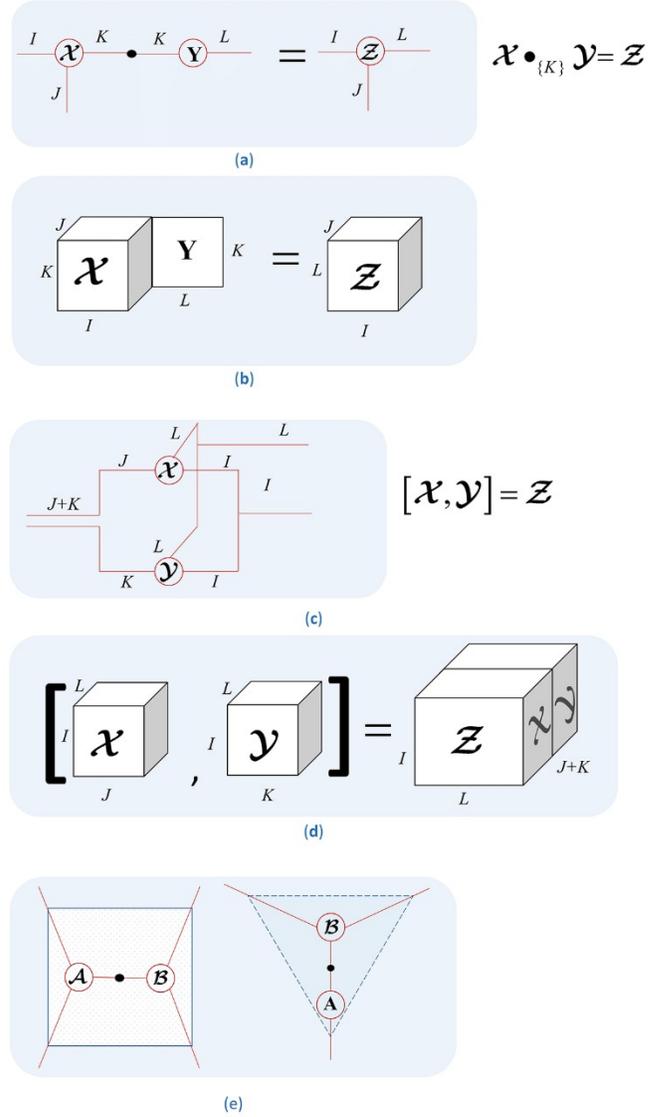

**Fig. 3.** Tensor operations with Penrose (P) Diagrams. (a,b) Graphical definitions of the contraction operator of $\mathcal{X} \in \mathbb{R}^{I \times J \times K}$ with $\mathcal{Y} \in \mathbb{R}^{K \times L \times M}$ giving $\mathcal{Z} \in \mathbb{R}^{I \times J \times L \times M}$. This operation is denoted with a black dot in a junction between the lines representing dimension $K$ on both tensors. See section II-A for a detailed formulation. (a) is the P diagram, (b) is the three dimensional representation of contraction of the same tensors. (c,d) Graphical definitions of the tensor concatenation operator is exemplified between two tensors of identical number of dimensions and elements per dimension with at most one mode being different, is this case $\mathcal{X} \in \mathbb{R}^{I \times J \times K}$ and $\mathcal{Y} \in \mathbb{R}^{I \times L \times K}$ gives $\mathcal{Z} \in \mathbb{R}^{I \times (J+L) \times M}$. As can be noted, the resulting tensor preserves all the dimensions of the original except the differing dimensions which is augmented with their sum. See Section II-A for a detailed formulation. (c) is the P diagram, (d) is the three dimensional representation of concatenation of the same tensors. (e) Tensors resulting from an operation involving other tensors can be surrounded by shapes (e.g triangles, squares) to highlight the resulting tensor, rather than on its parts.

## B. Markov-Penrose Diagrams

Missing from the usual P diagrams are the concepts of probabilistic dependency. It is, probabilistic graphical models are extremely useful for making explicit the conditional dependence in a statistical model [67], in particular directed graphical models (DAGs) or Bayesian networks for graphical models.

A junction between the links of two tensors in a P diagram signifies undirected arithmetic operations (as in Fig 3(a)). By contrast, an arrow between two variable nodes in a DAG signifies directed conditional dependence.

To specify statistical models for tensors we need both types of links. To summarize, P-diagrams are useful for the visualization of tensors and tensorial operations whereas DAGs are useful for showing the probability models.

This has prompted us to define a notation that incorporates both types of links as shown in Fig. 4(a,b,c). Here a directed arrow denotes a conditional probabilistic dependency between tensors.

Functional penalties $\pi(\mathbf{x})$ on nodes (such as those described in Section III) are depicted inside a special box shape node from which an arrow is directed to the constrained node. Note that this implies that the node variable is distributed as $p(\pi(\mathbf{x})) = C \exp(-\pi(\mathbf{x}))$ (see Fig. 4 (c)).

To contrast DAG with M-P diagram, we can now revisit the forward matrix based models introduced for the EEG and fMRI in Section III. Their respective M-P diagrams are shown in Fig. 4(d,e).

The DAG notation is operation blind, we can see the dependence between variables, but not the actual arithmetic. In M-P notation we can show both the same conditional dependence structure as in DAG notation, but we can also infer the formula for the model in this graph using the Penrose notation for operations.

There are other types of decompositions that might be applied to these models, such as Tucker decompositions [36] that are easily shown with this notation but, for reasons of space, will not be included in this review.

Note the inclusion in some DAG type arrow of an error term $\varepsilon$ indicates measurement noise. Of course in reality these can be drawn from any probability distribution but for ease of exposition, as in the matrix case, will be considered a tensor of adequate dimensions containing identically and independent Gaussian variates.

The close formal affinity between certain types of *tensor networks* and graphical models studied in the machine learning community has already been pointed out in [41] (see Table II of that reference) [68], [69] also highlight the similarities between particular types of tensor networks (depicting matrix product states) with Hidden Markov Models. Nevertheless, the links in these two types of structures are of a fundamentally different mathematical nature and a distinctive notation was not considered.

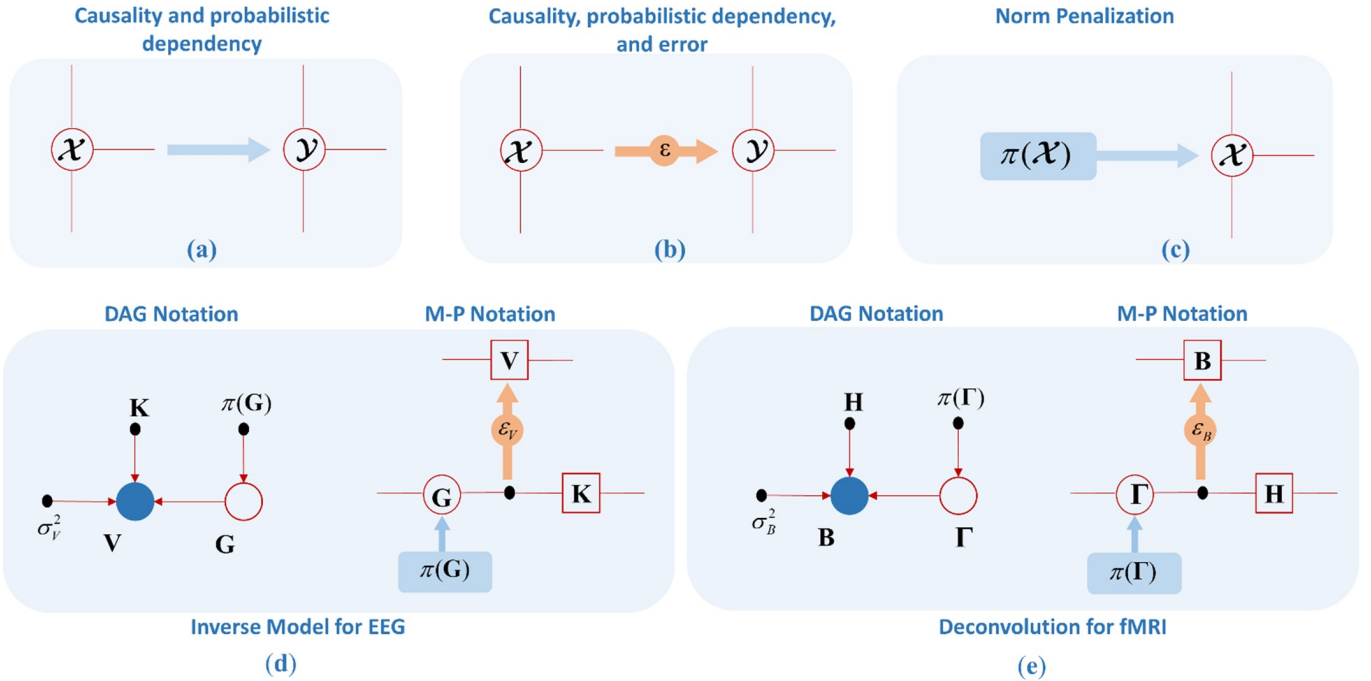

**Fig. 4.** Markov-Penrose (M-P) Diagrams (a) An arrow between two tensors denotes probabilistic causality between them, (b) if we add a $\varepsilon$ circle to the arrow, we indicate additive error term, (c) if the emitter of the arrow is a probabilistic density function this denotes a prior distribution for the variable it is pointing to. This is quite similar to the usual Directed Acyclic Graph (DAG) notation. (d) Usual DAG [67] diagrams for the generative models of the EEG (left) and its equivalent version in the M-P notation (right). The M-P diagram explicitly states that the EEG **V** is the contraction of the lead field matrix **K** with the primary current density matrix **G** corrupted by sensor noise $\varepsilon$. This example model also includes a prior probability distribution for **G**. (e) Similar DAG and M-P diagrams for the generative model of the BOLD signal. Here the vasoactive feedforward signal **Γ** that is contracted with the hemodynamic response function **H** (temporal convolution) producing the BOLD signature **B**. A prior for **B** is also shown.

## V. TENSOR EEG ANALYSIS

### A. Parallel Factor Analysis of scalp EEG

One of the main advantages of tensor based analyses is the ability to represent large multidimensional arrays in terms of much simpler structures. The best known of these representations is the Canonical Decomposition or Parallel Factor Analysis (PARAFAC) decomposition, a tensor equivalent to Principal Component Analysis. As mentioned before, for tensors of order equal or larger than 3 this decomposition is unique under rather mild conditions [31], [36].

For the sake of concreteness we shall illustrate PARAFAC in the context of time/frequency decompositions of the EEG which naturally lead to three way tensors. Calculating the wavelet or Gabor spectrum for each channel $\mathbf{V}(i,:)$ $i=1...I_E$ yields a matrix $\mathcal{S}_T(i,:,:)$ for given time frames and frequencies. The spectra for all channels may be shaped into a three way tensor $\mathcal{S}_T$ of size $I_E \times I_{T\delta} \times I_{F\delta}$ where $I_E$ is the number of sensors, $I_{T\delta}$ is the number time samples and $I_{F\delta}$ is the number of frequency samples.

The PARAFAC model decomposes $\mathcal{S}_T$ into $R$ components or "atoms" which can be expressed in several alternative forms. In terms of its scalar components it is:

$$\mathcal{V}_T(i_E, i_{T\delta}, i_{F\delta}) = \sum_{r=1}^{R} \mathbf{M}_\mathbf{V}(i_E, r) \mathbf{T}_\mathbf{V}(i_{T\delta}, r) \mathbf{F}_\mathbf{V}(i_{F\delta}, r) + \mathcal{E}(i_E, i_{T\delta}, i_{F\delta}) \quad (8)$$

where $i_E$, $i_{T\delta}$ and $i_{F\delta}$ are indices for space, time and frequency respectively and $\mathcal{E}$ denotes noise. $\mathbf{M}_\mathbf{V}(:,r)$, $\mathbf{T}_\mathbf{V}(:,r)$, $\mathbf{F}_\mathbf{V}(:,r)$ are the spatial, temporal and spectral signatures respectively for atom $r$.

To make clear the M-P notation, we re-express (8) in terms of tensor operations.

$$\mathcal{S}_T = \mathbf{M}_\mathbf{V} \bullet_{\{R\}} \mathbf{T}_\mathbf{V} \bullet_{\{R\}} \mathbf{F}_\mathbf{V} + \mathcal{E}$$
$$[\![\mathbf{M}_\mathbf{V}, \mathbf{T}_\mathbf{V}, \mathbf{F}_\mathbf{V}]\!] + \mathcal{E} \quad (9)$$

Many multi-linear models, in particular PARAFAC, components have a scale indeterminacy which does not affect interpretation. By convention we shall assume that factors are all normalized except the first one in the Kruskal operation.

This is the model applied in [43] which isolated several rhythmic components of the EEG. There are several variations on this basic PARAFAC model that can improve interpretability. For example we may impose sparseness, smoothness, non-negativity and orthogonality for the spatial signatures similar to that of (3).

The resulting model - a PARAFAC/ICA - is depicted in M-P format as in Fig. 5. In practical terms this model may be estimated as follows:

$$(\hat{\mathbf{M}}_\mathbf{V}, \hat{\mathbf{T}}_\mathbf{V}, \hat{\mathbf{F}}_\mathbf{V}) = \arg\min_{\mathbf{M}_\mathbf{V}, \mathbf{T}_\mathbf{V}, \mathbf{F}_\mathbf{V}} \left\{ \frac{1}{2} \|\mathcal{S}_T - [\![\mathbf{M}_\mathbf{V}, \mathbf{T}_\mathbf{V}, \mathbf{F}_\mathbf{V}]\!] \|_2^2 + \lambda_1 \|\mathbf{M}_\mathbf{V}\|_1 + \lambda_2 \|\mathbf{L}\mathbf{M}_\mathbf{V}\|_2^2 \right\} \quad (10)$$
$$\text{s.t.} \, \mathbf{M}_\mathbf{V}^T \mathbf{M}_\mathbf{V} = \mathbf{I}, \mathbf{M}_\mathbf{V} \geq 0, \mathbf{F}_\mathbf{V} \geq 0$$

The M-P diagram of this model is presented in Fig. 5. As will be seen in the next subsection this model is easily combined with EEG source estimation.

A review of related work on combining tensor methods for EEG analysis may be found in [70]. The combination of ICA with PARAFAC has also been developed by [71], [72]. A slightly different approach - applied to fMRI data - in [73] is to impose the PARAFAC structure during ICA extraction.

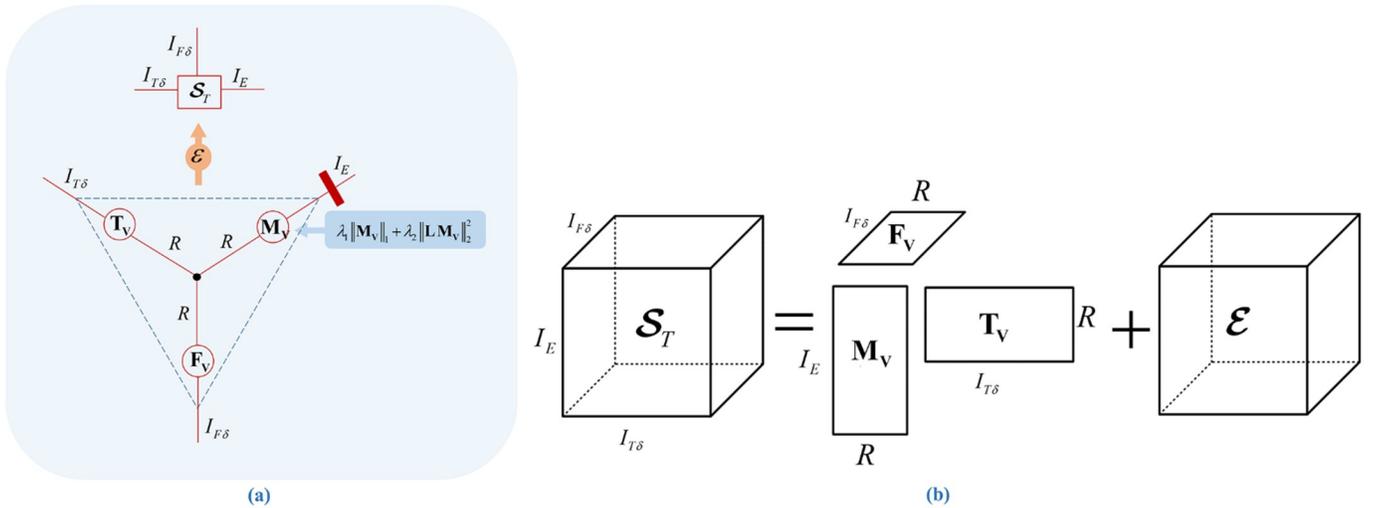

**Fig. 5.** (a) PARAFAC/ICA decomposition for a third order tensor, $\mathcal{V}_T$. Atoms (components) are latent variables shown with circles and $\mathcal{V}_T$ is the observed variable. Here $\mathbf{M}_\mathbf{V}$ is required to be orthogonal, nonnegative and to be sparse and smooth. (b) The more traditional 3D dimensional representation of PARAFAC.

A characteristic of the PARAFAC model that underlies our presentation, is that, as a consequence of the outer product structure of signatures, reconstructed atoms may appear as "oval blobs". While useful as a first approximation, this is a limitation that may be taken care of by including non-linear crossed terms as variables or by using more complex multi-linear models (e.g. Tucker decomposition). This is a topic that will be the subject of future research.

*B. Tensor Based EEG Inverse Problems*

In Section III-A we described STONNICA, a type of ICA in source space that has applied to time domain electrophysiological data. We have already seen that EEG time/frequency decompositions are best described in a tensor format and that PARAFAC decompositions may reveal interesting latent structures. It is therefore natural to integrate PARAFAC and STONNICA, a model that is estimated as follows:

$$\mathcal{S}_T = [\![ \mathbf{K}\mathbf{M}_G, \mathbf{T}_V, \mathbf{F}_V ]\!]$$

For which the estimator will be:

$$(\hat{\mathbf{M}}_G, \hat{\mathbf{T}}_V, \hat{\mathbf{F}}_V) = \arg\min_{\mathbf{M}_G, \mathbf{T}_V, \mathbf{F}_V} \left\{ \begin{array}{l} \frac{1}{2} \| \mathcal{S}_T - [\![\mathbf{K}\mathbf{M}_G, \mathbf{T}_V, \mathbf{F}_V]\!] \|_2^2 + \\ \lambda_1 \|\mathbf{M}_G\|_1 + \frac{1}{2}\lambda_2 \|\mathbf{L}\mathbf{M}_G\|^2 \end{array} \right\} \quad (11)$$

$$\text{s.t. } \mathbf{M}_G^T \mathbf{M}_G = \mathbf{I}, \mathbf{M}_G \geq 0, \mathbf{F}_V \geq 0$$

$\mathbf{F}_V$ is the spectral signature of the generator sources of the EEG. The Markov Penrose diagram of this model is shown in Fig 6(a).

The application of this model is to the analysis of the resting state EEG is analysis shown in Fig. 6(b). Note that three atoms were identified, all predominant in the occipital cortex. The frequency signatures are shown in the top left of that figure, showing that the components were primarily related to the different types of alpha rhythm.

Related work on combining PARAFAC and inverse solutions can be found in [74].

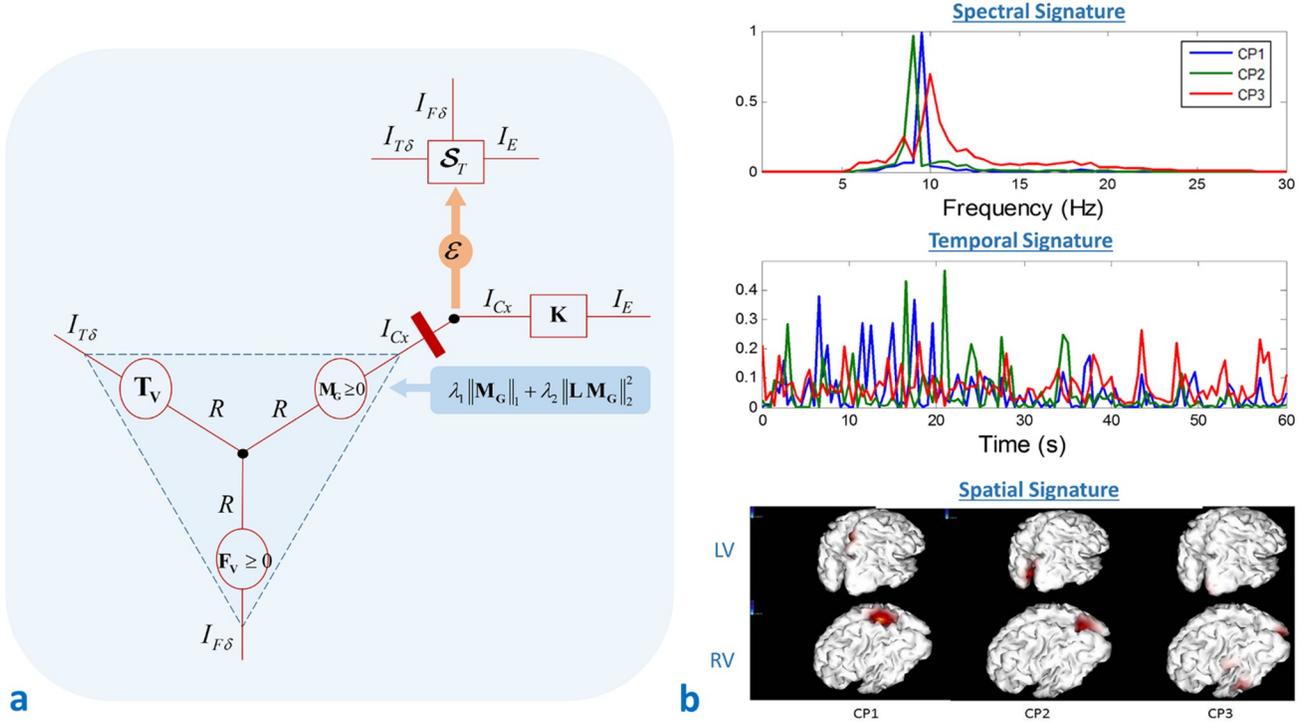

**Fig. 6.** (a) M-P Diagram indicating the tensor operation flow for tensor STONNICA. In this case a PARAFAC decomposition of an inverse solution, being $\mathcal{S}_T$ the signal observed in the scalp, $\mathbf{K}$ the lead field matrix. The atoms of the decomposition are denoted as spectral signature ($\mathbf{F}_V$ in the diagram), temporal signature ($\mathbf{T}_V$) and spatial localization of sources ($\mathbf{M}_G$), for the latter of which multiple regularizations are imposed. (b) An example of the application of this procedure on the resting state EEG from one subject. Tensor STONNICA is applied on the cross-spectral density of EEG for all segments in the frequency range 0.5 Hz – 30Hz in 0.5 Hz steps. MNI based head model is used for the calculation of lead field. Spectral signatures show three distinct atoms at 9, 9.5 and 10 Hz all of them being alpha atoms. Temporal signatures reveal co-existence of these rhythms at the same time at different magnitudes. Spatial signatures in source space are localized in occipital, striatal and para-striatal areas. Note that, three atoms are well separated in frequency and spatial localization (For more information on the dataset refer to [139]).

## VI. BRAIN CONNECTIVITY AS A TENSOR REGRESSION

We now address the use of tensor methods to evaluate brain connectivity. Of the different types shown in Figure 7 we shall concentrate on **effective connectivity.** This is the direct causal activation of one neural mass by another mediated by axonal pathways.

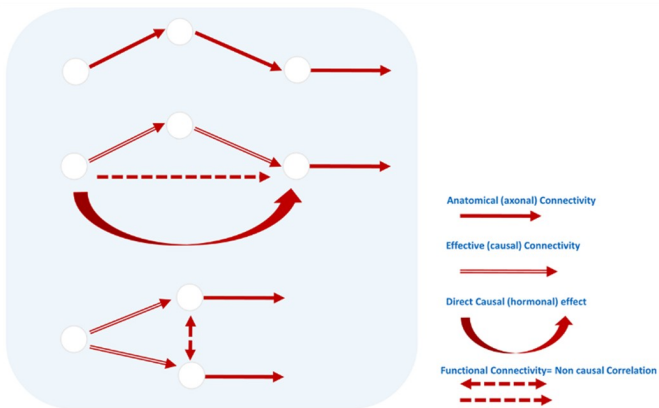

**Fig. 7.** Different types of Brain Connectivity. Several related concepts of brain connectivity are summarized. Anatomical connectivity is defined as the existence of axonal pathways that link two distinct neural masses. If this pathway is activated there is effective connectivity between these neural masses. On the other hand if one only measures a correlation between the activities of two neural masses one is measuring functional connectivity. It is effective connectivity that is of interest in defining functional neural networks.

This topic has been reviewed in in depth in [75]. In that paper it is argued that current work in this area is a fusion of several strands of research:

- One strand is based on graphical representations of causal models and the conditions under which causal inference is possible. This **structural** approach is best exemplified by the work of Pearl [76] who formalizes role of interventions to determine true causal relations.
- A second line of work is that of Wiener [77], Akaike [78], Granger [79] and Schweder [80] in which predictability of one time series by another is used to define Granger (actually **WAGS**) influence measures.
- A third strand is the integration of WAGS theory with the structural approach previously mentioned which has been carried out for discrete time by Eichler [81] and for continuous time [82].
- Structural and WAGS influence are amplified by the **biophysical** modeling (c.f. by random differential equations) as best illustrated by Dynamic Causal Modeling [83].

Biophysical causality explicitly examines the state space equations that not only model system dynamics but also the observation equation [75]. From the discussion of the EEG forward problem in Section III-A, it becomes evident that volume conduction is a serious problem affecting the interpretation of influence measures. For a recent attempt to overcome the complexities of EEG Granger causality see [84].

Since our purpose is to show that the estimation of Granger Causality (influence) measures for fMRI data can be profitably approached using tensor methods, we exclude the EEG from this section and will limit our examples on a standard dataset of BOLD measurements.

This is fast fMRI data (sampled at 10 Hz) from one of the subjects reported in [85]. The subject recorded had to respond with the corresponding hand to right or left visual hemi-field stimuli. Data from 1100 voxels were recorded from the visual (V), parietal (PCC), premotor (PreM), somatosensory (S), and motor (M) regions of interest (ROIs).

### A. WAGS influence or Granger Causality (GC)

This influence measure is based upon the Multivariate Autoregressive model (MAR). For a review and freely available toolbox see [86]. The algorithms described in this and many other papers are useful only when analyzing a quite small number of time series. We first set out its matrix formulation.

As pointed out in [75], [87], for brain imaging data a high dimensional MAR is needed to search for the influence fields that are the spatial maps of the influence of one brain area on rest of the brain.

To formalize these ideas, we remind the reader that the BOLD signal is denoted by $\mathbf{B} \in \mathbb{R}^{I_{Cx} \times I_{T\delta}}$. Then the spatial MAR is:

$$\mathbf{b}_t = \sum_{i_{lag}=1}^{I_{lag}} \mathbf{A}_{i_{lag}} \mathbf{b}_{t-i_{lag}} + \mathbf{e}_t \tag{12}$$

where $\mathbf{b}_t = \mathbf{B}(:,t)$, $I_{lag}$ is the number of past lags included in the model and $\mathbf{e}_t$ is the innovation noise.

The autoregressive matrices $\mathbf{A}_{i_{lag}} \in \mathbb{R}^{I_{Cx} \times I_{Cx}}$ quantify the influence of the past of a time series on all others including itself. If a coefficient $\mathbf{A}_{i_{lag}}(i,j) \neq 0$ we shall say that time series $j$ (Granger) influences time series $i$ after $i_{lag}$ lags.

Thus GC measures are essentially tests of the null hypothesis for coefficients of the MAR model.

One of the main problems is that for fMRI the number of time series $I_{Cx}$ is much larger than the length of the time series $I_T$. A MAR model contains $I_{lag} \cdot I_{Cx}^2 + (I_{Cx}^2 + I_{Cx})/2$ variables. This reveals (12) as a high dimensional $p \gg n$ regression problem for which the usual multivariate statistical techniques fail.

To avoid the high dimensional scenario, the problem can be reduced in size, first by averaging the BOLD signal over voxels for pre-selected ROI, and then carrying out bivariate GC for all pairs of ROI—correcting for multiple comparisons. The results of this analysis for our standard dataset is shown in Fig. 10(a) in which an outflow of influence from the VC spreads out to other ROIs.

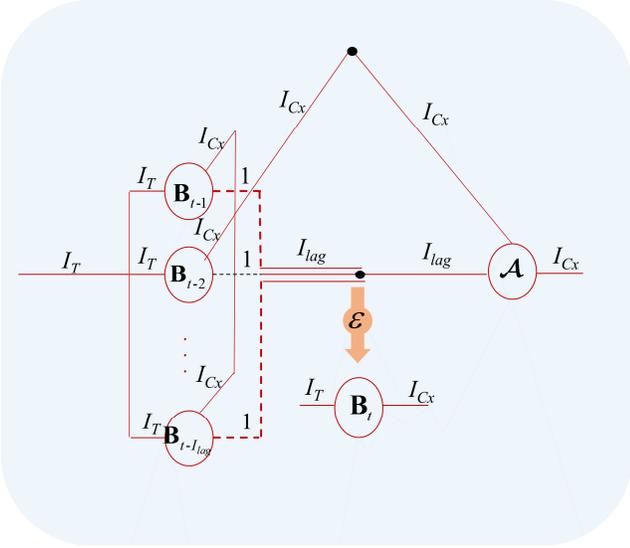

**Fig. 8.** Representation of Granger Causality as a tensor regression problem. The time series $\mathbf{B}_{t-i_{lag}}$ of size $I_{Cx} \times I_T$ are constituted by lagging a sample matrix $\mathbf{B}_t$ at $i_{lag} = 1, \ldots, I_{lag}$. Concatenation of the $\mathbf{B}_{t-i_{lag}}$'s over $I_{lag}$ singleton dimensions gives the lagged data tensor $\mathcal{B}$ with size $I_{lag} \times I_{Cx} \times I_T$. $\mathcal{B}$ is contracted with connectivity tensor $\mathcal{A}$ over spatial and temporal lag dimensions resulting the time-series in voxels $\mathbf{B}_t$.

An alternative approach is to use methods for high dimensional data regression [88]. This penalized MAR approach was first presented in [87] and reviewed in [47]. Here all voxels of interest (the whole brain if necessary) may be included in the regression but statistical procedures that lead to variable selection were carried out. However only matrix based models were dealt with.

We now change our viewpoint and look at the penalized MAR model in tensor format.

### B. Granger Causality viewed as a Tensor Regression

If we collect the $\mathbf{b}_t$ from (12) for all time samples $t = I_{lag} + 1, \cdots, I_T + I_{lag}$, we define a matrix $\mathbf{B}_{t-i_{lag}} \in \mathbb{R}^{I_{Cx} \times I_T}$ as follows:

$$\mathbf{B}_{t-i_{lag}} = \left[\mathbf{b}_{I_{lag}+1-i_{lag}}, \cdots, \mathbf{b}_{I_T + I_{lag} - i_{lag}}\right]^T$$

Then we concatenate the matrices $\mathbf{B}_{t-i_{lag}} \in \mathbb{R}^{I_{Cx} \times I_{T\delta}}$ constructed for all $i_{lag} = 1, \cdots, I_{lag}$ to obtain a data tensor $\mathcal{B} \in \mathbb{R}^{I_{lag} \times I_{Cx} \times I_T}$.

The autoregressive coefficients are also essentially a tensor $\mathcal{A} \in \mathbb{R}^{I_{Cx} \times I_{Cx} \times I_{lag}}$ that is obtained by concatenating $\mathbf{A}_{i_{lag}}$ matrices along the $I_{lag}$ dimension. We state the concatenation operations explicitly as follows:

$$\mathcal{B} = [\mathbf{B}_{t-1}, \ldots, \mathbf{B}_{t-i_{lag}}, \ldots, \mathbf{B}_{t-I_{lag}}]$$
$$\mathcal{A} = [\mathbf{A}_1, \ldots, \mathbf{A}_{i_{lag}}, \ldots, \mathbf{A}_{I_{lag}}]$$

Note that we make use of the property of tensors that adding singleton dimensions to a tensor will not change the dimension. The concatenation operation for $\mathcal{B}$ is depicted in Fig. 8.

Thus the MAR model is a tensor regression expressed by means of a tensor contraction:

$$\mathbf{B}_t = \mathcal{A} \bullet_{\{I_{Cx}, I_{lag}\}} \mathcal{B} + \varepsilon \tag{13}$$

The M-P diagram for this type of model is given in Fig. 8. Note that $\mathcal{A} \in \mathbb{R}^{I_{Cx} \times I_{Cx} \times I_{T\delta}}$ is a three dimensional tensor in the same figure.

In recognition of the tensor nature of the MAR in (13) we propose following estimation procedure:

$$\hat{\mathcal{A}} = \arg\min_{\mathcal{A}} \left\{ \left\| \mathbf{B}_t - \mathcal{A} \bullet_{\{I_{Cx}, I_{lag}\}} \mathcal{B} \right\|_2^2 + \pi(\mathcal{A}) \right\} \tag{14}$$

--For example in [47] the penalty $\pi(\mathcal{A}) = \lambda_1 \|\mathcal{A}\|_1 + \lambda_2 \left\| \mathcal{L} \bullet_{\{I_{Cx}\}} \mathcal{A} \right\|^2$ was used, as well as a number of other variants.

### C. Granger Causality with t-Products

An alternative is to use tensor norms as penalization functions. Such an approach takes advantage of the *t-product* to implement the Levinson-Durbin estimation of MAR [89].

We need the following definitions:

- The sample covariance tensor $\mathcal{R} \in \mathbb{R}^{I_{Cx} \times I_{Cx} \times (I_{lag}+1)}$ is
$$\mathcal{R}(i_{Cx}, i_{Cx}, i_{lag}) = \frac{1}{I_T} \sum_{i_T=1}^{I_T} \mathcal{B}(i_{lag}, i_{Cx}, i_T) \mathbf{B}_t(i_{Cx}, i_T),$$ where each block $\mathcal{R}(:,:,i_{lag})$ is an $I_{Cx} \times I_{Cx}$ cross-covariance matrix.

- $\mathcal{R}_1 = \mathcal{R}(:,:,0:I_{lag}-1); \quad \mathcal{R}_2 = \mathcal{R}(:,:,1:I_{lag});$

The naïve solution to the classical Levinson-Durbin equation in our notation is:

$$\text{MatVec}(\mathcal{A}) = \text{tplz}(\mathcal{R}_1)^{-1} \text{MatVec}(\mathcal{R}_2) \tag{15}$$

It is well known that this type of solution is not numerically stable. Therefore one approach is to regularize the estimate of the covariance matrix $\mathcal{R}_1$ using the t-operations defined in Section II-B. The specific estimator is:

$$\hat{\mathcal{R}}_1 = \arg\min_{\Lambda} \left\{ \left\| \mathcal{R}_1 - \Lambda \right\|_2^2 + \lambda \|\Lambda\|_{\circledast} \right\} \tag{16}$$

where the penalty term is the of tensor nuclear norm (TNN) defined in Section II-B.

The estimator $\hat{\mathcal{R}}_1$ can be explicitly found by shrinking the t-singular values in $\mathcal{D}$ by $\lambda$ by applying the $\rho$ function of [90]:

$$\hat{\mathcal{R}}_1 = \mathcal{U} *_t \rho(\mathcal{D}) *_t \mathcal{V}^T$$

Then $\mathcal{A}$ is estimated as:

$$\hat{\mathcal{A}} = \mathcal{V} *_t \rho(\mathcal{D}) *_t \mathcal{U}^T *_t \mathcal{R}_2$$

In fact this operation was improved by using the circulant embedding defined in [91].

This model was applied to the standard dataset and the resulting connectivity diagram is presented in Fig 10(b). The method was able to deal high dimension, with more than 1000 nodes and 20 lags with stable results numerical results. It is also interesting to note that this estimate of connectivity seems to be much more sensitive than the simple bivariate approach.

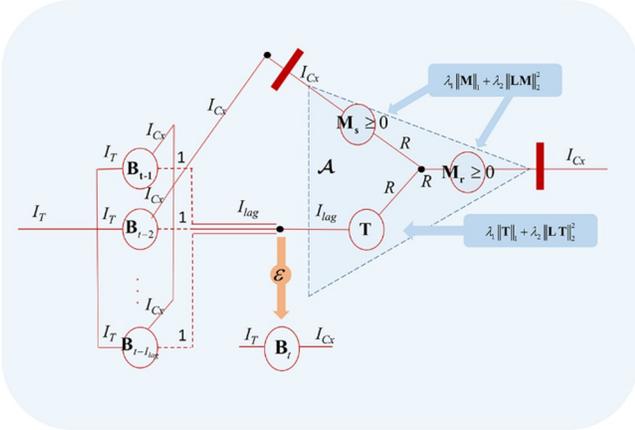

**Fig. 9.** Granger Causality as a PARAFAC decomposition. The connectivity tensor indicating voxel to voxel causal effect (denoted as tensor $\mathcal{A}$ in Fig. 8) is decomposed by PARAFAC of order $R$ into three component matrices: $\mathbf{M}_r$ for receiver nodes, $\mathbf{M}_t$ for transmitter nodes and $\mathbf{T}$ for temporal lags. For the sake of interpretation, multiple priors are used for the estimation of components: sparsity, smoothness, orthogonality for spatial factors, and sparsity and smoothness for temporal lags. Since the connectivity tensor may take negative values, the temporal factor is kept to be real valued with the purpose of sign convention.

*D. Granger Causality with PARAFAC/ICA*

As seen the tensor methods uncover a much richer set of connections than simpler methods. However basic neuroscience suggests that effective connectivity implies a structured sparsity of $\mathcal{A}$, which is desirable since it would prune many of the spurious connections characteristic of some functional connectivity measures. Structured sparsity can be achieved by positing a PARAFAC/ICA structure for the connectivity tensor. We shall define a node as a *sender* if it influences another set of nodes, and *receiver* if its activity is caused by other nodes.

Signatures of $\mathcal{A}$ are estimated by:

$$(\hat{\mathbf{M}}_s, \hat{\mathbf{M}}_r, \hat{\mathbf{T}}) = \underset{\mathbf{M}_s, \mathbf{M}_r, \mathbf{T}}{\arg\min} \begin{Bmatrix} \frac{1}{2}\|\mathbf{B}_t - \mathcal{A} \bullet_{\{I_{Cx}, I_{lag}\}} \mathcal{B}\|_2^2 \\ + \lambda_1 \|\mathbf{M}_s\|_1 + \frac{1}{2}\lambda_2 \|\mathbf{L}\mathbf{M}_s\|^2 \\ + \lambda_3 \|\mathbf{M}_r\|_1 + \frac{1}{2}\lambda_4 \|\mathbf{L}\mathbf{M}_r\|^2 \\ + \lambda_5 \|\mathbf{T}\|_1 + \frac{1}{2}\lambda_6 \|\mathbf{L}\mathbf{T}\|^2 \end{Bmatrix}$$
$$\text{s. t. } \mathcal{A} = [\![\mathbf{M}_s, \mathbf{M}_r, \mathbf{T}]\!], \mathbf{M}_s \geq 0, \quad (17)$$
$$\mathbf{M}_s^T \mathbf{M}_s = \mathbf{I}, \mathbf{M}_r \geq 0, \mathbf{M}_r^T \mathbf{M}_r = \mathbf{I}$$

where $\mathbf{M}_r$ is the spatial signature for *receiving* nodes, $\mathbf{M}_s$, is the spatial signature for *sender* nodes, and $\mathbf{T}$ the temporal signature for causal lags.

In this model, the identifiability is enhanced by enforcing nonnegativity, orthogonality, smoothness and sparseness for the spatial signatures and a smooth lasso type constraint for the lag signature. In other words these constraints tend to estimate smooth patches of voxels on the cortex. Orthogonality and nonnegativity constraints guarantee that spatial factors can have only one nonnegative element in each row which can be interpreted as the cluster centroids [92], [93]. In this way the connected spatial regions are confined to be non-overlapping patches. This model is the generalization of clustering in which connectivity tensor is decomposed into sum of rank one tri-clusters [94].

Atomic decomposition of tridimensional connectivity tensor for the model of (17) favors a parsimonious model where the number of parameters to be estimated is $(2I_{Cx} + I_{lag})R$, with $R$ being the model order of PARAFAC. M-P diagram is shown in Fig. 9.

For the application GC-PARAFAC on the fMRI standard dataset, a time period of 500 milliseconds corresponding to 5 time frame lags was selected as the temporal factor. A graph Laplacian matrix is used as the smoother matrix $\mathbf{L}$ in (17). The model order of PARAFAC was set to 3. Fig. 10(c) shows the existence of strong bottom-up and weak top-down connections between VC, PCC, M and S. There is also lateral information flow from left to right visual areas.

We wish to note that the work in this section was encouraged by the tensor formulation of the State Space Model set out by [95]. However this is, to our knowledge, the first time the estimation of the well-known MAR model has been posed as a tensor regression problem. As we have just seen this has suggested interesting uses of tensor methods for analyzing WAGS influence for fMRI signals.

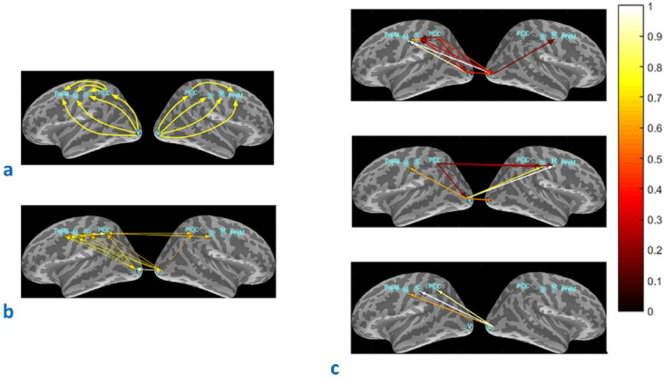

**Fig. 10.** Granger Causality on Real Data. The data corpus described in Section VI was used for analyzing brain connectivity by means of different techniques, which offer comparable results. The arrows denote directional dominant flows of Granger causality between the visual (**V**), parietal (**PPC**), premotor (**PreM**), somatosensory (**S**), and motor (**M**) cortex. (a) The original results were published by [85] and extracted from Fig 2 in that paper. This is the dominant information flow calculated from the difference between two uni-directional Granger estimates among the ROIs. Only connections that have a p-value $\geq 0.05$ are shown. (b) Results using the t-product technique described in Section VI make more valuable the connectivity analysis revealing the bi-directional and inter hemispheric interaction between ROIs. A major interaction was found at the left hemisphere that could be related to the right handed condition of the subject. (c) The resulting three spatial atoms retrieved by the PARAFAC decomposition of the connectivity matrix. Note that in all of these results there is a predominance in causal directionality emerging from the **V** and **PPC** cortex to the rest of the areas. Magnitude of the connectivity is symbolized by the color bar at the right of figure.

## VII. TENSOR EEG/fMRI FUSION

As reviewed in [23] symmetrical fusion of EEG/fMRI is an equal opportunity combination of modalities in which the influence of each modality to the final solution is selected by the data. Fusion can be of two types: model driven and data driven. We will concentrate on data driven approaches as an extension of the matrix based approaches of Section II-C. We recall Fig 1 and note that we will assume that the VFFS $\mathbf{\Gamma}$ and the primary current density $\mathbf{G}$ related to each other in a simple fashion.

Fusion can be carried out by using tensor techniques that link the two modalities along a common dimension, usually temporal or spatial. Another possibility for multi-subject data is to use subject identity as the common link. In this section we will give three examples of fusion: EEG/fMRI along the temporal domain, EEG/DTI along the subject domain and EEG/fMRI along the spatial domain. We now explain these examples.

### A. Multiway Partial Least Squares (N-PLS)

Multiway partial least squares is a tensor based method in which both dependent and independent variables constituted as tensors are decomposed simultaneously, while the signatures or factors of the shared dimension are required to have maximal covariance. In [45] to find the BOLD correlates of EEG rhythms, time-varying EEG spectrum $\mathcal{S}_T$ is decomposed into spatial $\mathbf{M_V}$, temporal $\mathbf{T_V}$ and spectral $\mathbf{F_V}$ signatures and fMRI data matrix $\mathbf{B}$ is decomposed into spatial $\mathbf{M_B}$ and temporal $\mathbf{T_B}$ signatures such that temporal factors $\mathbf{T_V}$ and $\mathbf{T_B}$ will have the maximum covariance. We therefore set the regression model between temporal signatures by:

$$\begin{aligned}\mathbf{T_B} &= \mathbf{T_V}\mathbf{C} + \varepsilon_t \\ \text{s.t.}\ \mathcal{S}_T &= [\![\mathbf{M_V}, \mathbf{T_V}, \mathbf{F_V}]\!] \text{ and } \mathbf{B} = [\![\mathbf{M_B}, \mathbf{T_B}]\!]\end{aligned} \quad (18)$$

Source localization is performed on the corresponding spatial signatures of EEG ($\mathbf{M_V}$) to find the generators of each component. Then the source signatures of EEG denoted by $\mathbf{M_G}$ is estimated through the forward model defined by:

$$\mathbf{M_V} = \mathbf{K}\mathbf{M_G} + \varepsilon_{M_V} \quad (19)$$

The M-P diagram of this type analysis is shown in Fig. 11.

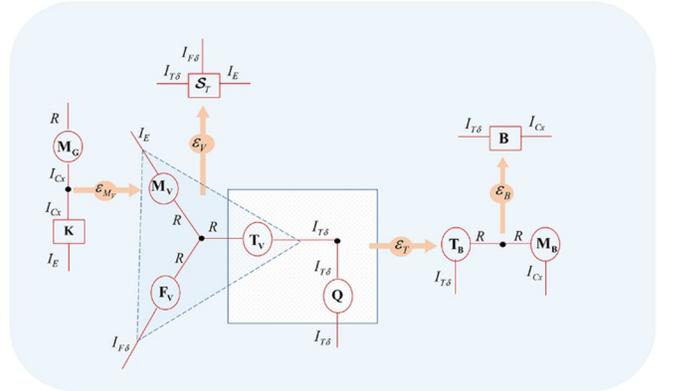

**Fig. 11.** M-P diagram of N-PLS for EEG/fMRI fusion. The EEG data $\mathcal{S}_T$ is decomposed using an $R$ order PARAFAC into temporal ($\mathbf{T_V}$), spectral ($\mathbf{F_V}$) and spatial $\mathbf{M_V}$ atoms, the latter of which is complemented by the source localization performed via the lead field matrix $\mathbf{K}$, resulting in a spatial cortical atom $\mathbf{M_G}$. A similar $R$ order decomposition is done on the fMRI data $\mathbf{B}$, with temporal ($\mathbf{T_B}$) and spatial ($\mathbf{M_B}$) atoms in a way that covariance of temporal factors $\mathbf{T_V}, \mathbf{T_B}$ is maximized via the matrix $\mathbf{Q}$.

This method was applied on the resting state EEG/fMRI data of a single subject originally collected in [96]. Since this has been a well-studied dataset we will only briefly mention the results that can be found in detail in [45].

The time varying EEG spectra $\mathcal{S}_T \in \mathbb{R}^{I_E \times I_{T\delta} \times I_{F\delta}}$ is estimated for $I_E = 16$ channels in the frequency range of 0.5 to 50 Hz for $I_{F\delta} = 124$ frequency points and for $I_{T\delta} = 105$ time points. fMRI data is acquired in six adjacent slices that cut through occipital lobe and thalamus.

The N-PLS analysis described in (18) is applied by maximizing the covariance between temporal signatures. The scatterplot of the atoms identified shows this correspondence.

Both topographic and source spatial signatures of EEG, spatial signature of fMRI are shown in Fig.12.

Three atoms were extracted exhibiting spectral peaks in the alpha $\alpha$, theta $\theta$ and gamma $\gamma$ range of the EEG. There was a significant covariance between EEG and fMRI components only for the $\alpha$ and $\theta$, which are the only ones we will comment on. Note in Fig. 12 that both the scalp and source topographies are clearly delimited as being in the occipital region ($\alpha$) and the frontal lobes ($\theta$). The $\theta$ atom showed frontal negative activity means that increased BOLD signal corresponds to decreased spectral activity. Note that the spatial signature of the $\alpha$ atom is positive for the thalamus and negative for the visual cortex. This negative associations between BOLD and EEG activity in cortex have been interpreted as due to desynchronization of neural activity with greater thalamo-cortical input; a hypothesis that has been received support with a large scale neural model described in [23].

Identifying the coupling between EEG rhythms and resting state fMRI is a topic of current great interest. An example for this type of work is [97]. Thus it seems that tensor methods could be of use for the study of these inter related phenomena.

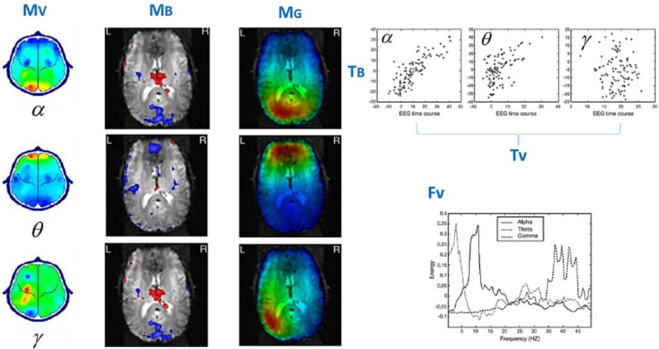

**Fig. 12.** Multiway Partial Least Squares results of EEG/fMRI Fusion. Topographic maps of the EEG ($\mathbf{M_V}$) and the corresponding source maps of the EEG ($\mathbf{M_G}$), spatial signature of fMRI ($\mathbf{M_B}$) are shown in columns. For all spatial maps red color indicates positive values and blue color indicates negative values. Scatter plot of the fMRI temporal signatures ($\mathbf{T_B}$) against EEG temporal signatures ($\mathbf{T_B}$) are demonstrated for three atoms. Spectral signature of EEG ($\mathbf{F_V}$) is also shown. Signatures are named as alpha ($\alpha$), theta ($\theta$) and gamma ($\gamma$) due to the resemblance of their spectral characteristics to EEG rhythms as revealed in the plot of the spectral signature of EEG. EEG spatial signatures show activation in occipital regions for alpha, frontal regions for theta and pario-temporal regions for gamma atoms. On the other hand for alpha atom, fMRI spatial signature shows negative activation in occipital and superior temporal regions and positive activation in the thalamus and insula. Theta atom of fMRI spatial signature shows negative activation in anterior cingulate and occipital regions. Gamma atom of fMRI spatial signature shows similar activity with the alpha atom. Scatter plot of fMRI temporal signature against EEG temporal signature significantly reveals positive correlation in alpha atom. Correlation of these signatures for theta and gamma atoms are found insignificant. EEG spectral signatures show three distinct peaks at alpha, theta and gamma frequencies. Note the topographically distinct pattern of relations BOLD signals and spectral components that indicates stable relations between fMRI, resting state components and EEG oscillations. Figure is adapted from [45].

A recent application of N-PLS to a different pair of modalities was to the joint decomposition of EEG and DTI functional anisotropy (FA) This was carried out in order to explore the neuroanatomical determinants of the inter-individual variability of the peak frequency of the EEG scalp alpha rhythm (8-12 Hz). Here the common dimension for both modalities was subject identification with an N of 200.

The data for each modality was encoded by an array with the common "subject dimension". The first modality is related to white matter architecture, as measured by DTI-FA. The voxels of the FA images in which white matter probability is lower than 0.5 and FA is below 0.1, were masked out, leaving $I_{Wm} = 24764$ voxels. The masked FA images were vectorised in rows and concatenated over subject dimension, $I_W = 200$ to build a two dimensional matrix $\mathbf{FA} \in \mathbb{Z}^{I_W \times I_{Wm}}$. Since $\mathbf{FA} \in [0,1]$ they were logit-transformed to approximate normal distribution. The second modality is EEG scalp spectrum estimate, organized in a tridimensional array $\mathcal{S}_S \in \mathbb{R}^{I_E \times I_W \times I_{F\delta}}$, being $I_E = 76$ the number of derivations (a subset of EEG channels of the 10-20 system) and $I_{F\delta} = 58$ is the number of frequencies in the range of 0.39-29.68 Hz. For the details on the MRI and EEG acquisition and preprocessing see [98]. Here both arrays were scaled and their grand mean subtracted. Inverse source localization was performed on the scalp spatial signatures of EEG using LORETA and the MNI-based head model.

For the correlation of EEG and FA data, N-PLS is used in which the covariance between subject signatures are maximized. We determined that 3 atoms were adequate for the description of the EEG dataset. Figure 13 shows the spectral signature $\mathbf{F_V}$, derivation signature $\mathbf{M_V}$, inverse source constructions $\mathbf{M_G}$ of the atoms obtained from the N-PLS.

The spectral and spatial signatures associated to the atoms allowed us to identify them as follows:

a) An atom with a spectrum dominated by "slow wave activity", with essentially a *1/f* decay. Its distribution was frontal. This atom might correspond with the EEG process identified as the "$\xi$ process" described in [99], [100].
b) An "alpha" ($\alpha$) atom with a spectral peak very similar to those shown for other in Figs 6 and 12. Its localization was occipital, as expected.
c) An "alpha contrast" atom, with a spectrum that resembles the derivate of that that $\alpha$ atom, also with an occipital topography.

It is interesting that the first two components are similar to those described for the $\xi - \alpha$ model [99], [100] which was obtained by individual parameterized spectral models for the EEG of a different set of 211 normal subjects.

The alpha contrast atom seems to reflect the individual to individual fluctuations of the peak alpha frequency across the sample. Interestingly, the subject signatures of both modalities for this atom have the highest and significant correlation ( $R = 0.7701, p \sim 10^{-8}$ ), suggesting that white matter architecture indeed is correlated with the inter-individual variation of the alpha rhythm. The spatial signature of the FA dataset $\mathbf{M}_{FA}$ for this atom, shown in Fig. 13 is loaded on the major tracts, especially thalamo-cortical, in accordance with the results in [98]. This type of result is important in order to test models of the origin of the alpha rhythm.

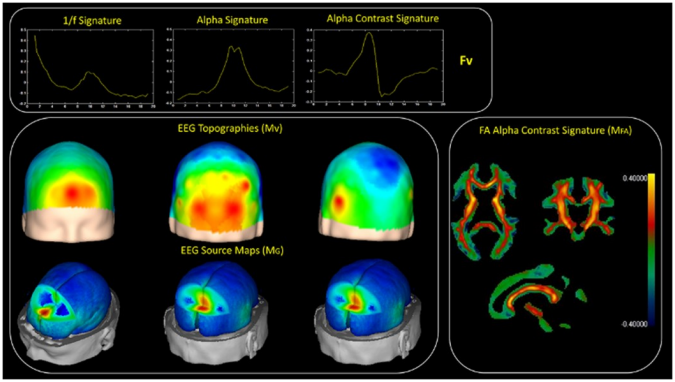

**Fig. 13.** Multiway Partial Least Squares for DTI-FA and EEG Fusion. Spectral signatures of EEG ($\mathbf{F}_V$) and spatial signature of FA ($\mathbf{M}_{FA}$) are shown. Both scalp distribution ($\mathbf{M}_V$) and their corresponding brain electrical tomography of the derivation signatures ($\mathbf{M}_G$) are demonstrated under the corresponding spectral signatures. The frequency signatures of the three EEG atoms show three distinct patterns. The first one is dominated by a low *1/f* behavior. We shall therefore identify this signature as the "slow wave signature". The second one is a pure alpha peak. We shall therefore identify this atom as the "alpha signature". The most interesting signature, the third one, seems to be a contrast between low and high alpha. We shall therefore identify this signature as the "alpha contrast signature". The slow wave signature shows a frontal source near to the theta signature described by [45]. As clearly seen the spectral alpha signature co-localizes with the source of the alpha rhythm as described by [45]. The new alpha contrast signature has a localization very similar to the alpha signature. We only showed the corresponding spatial signature of the FA to alpha contrast signature. Major tracts, especially talamo-cortical are enhanced in this signature in accordance with [98].

*B. Coupled Matrix-Tensor Factorization (CMTF)*

We propose a new data fusion framework based on a joint decomposition of EEG and fMRI along the common spatial profile. We extend the matrix based EEG/fMRI fusion in (7) to coupled tensor decompositions of $\mathcal{S}_T$ and fMRI data matrix **B**. In Section V- B we showed that source signatures of EEG can be identified from $\mathcal{S}_T$ by using PARAFAC. The EEG tensor $\mathcal{S}_T \in \mathbb{R}^{I_E \times I_{T\delta} \times I_{F\delta}}$ is decomposed into source spatial $\mathbf{M}_G$, temporal $\mathbf{T}_V$ and spectral $\mathbf{F}_V$ signatures, and the fMRI data matrix $\mathbf{B} \in \mathbb{R}^{I_{Cx} \times I_{T\delta}}$ is decomposed into spatial $\mathbf{M}_B$ and temporal $\mathbf{T}_B$ signatures in which source spatial signatures are coupled during decomposition.

Earlier fusion algorithms were predicated on the idea that the support of the EEG active regions had a complete coincidence with that of the fMRI. Unlike conventional CMTF algorithms where all dimension are considered to be common we distinguish also a discriminative subspace [101] where the signatures of both modalities are not overlapping. This enables us to deal with the cases in which EEG and fMRI sources may have a spatial mismatch [102]. Coupled and uncoupled spatial profiles are obtained for each modality.

Assume that $\mathbf{M}_{eeg}$ is the source spatial factor of $\mathcal{S}_T$ and $\mathbf{M}_{fmri}$ is the spatial factor of **B**, in the proposed framework these factors will be; $\mathbf{M}_{eeg} = [\mathbf{M}_C, \mathbf{M}_G]$ and $\mathbf{M}_{fmri} = [\mathbf{M}_C, \mathbf{M}_B]$ where subscript *C* is for common part and subscript *G* (*B*) is for discriminant factor of EEG (fMRI). In this way different model orders can be assigned to the decomposition of $\mathcal{S}_T$ and **B** as long as the number of common components are kept the same i.e. column number of $\mathbf{M}_C$.

Modality specific and coupled signatures are estimated by:

$$(\hat{\mathbf{M}}_C, \hat{\mathbf{M}}_G, \hat{\mathbf{T}}_V, \hat{\mathbf{F}}_V, \hat{\mathbf{M}}_B, \hat{\mathbf{T}}_B)$$
$$= \underset{\mathbf{M}_C, \mathbf{M}_G, \mathbf{T}_V, \mathbf{F}_V, \mathbf{M}_B, \mathbf{T}_B}{\arg\min} \begin{cases} \frac{1}{2} \| \mathcal{S}_T - [\![ \mathbf{K}[\mathbf{M}_C, \mathbf{M}_G], \mathbf{T}_V, \mathbf{F}_V ]\!] \|_2^2 \\ + \gamma \frac{1}{2} \| \mathbf{B} - [\![ [\mathbf{M}_C, \mathbf{M}_B], \mathbf{T}_B ]\!] \|_2^2 \end{cases} \quad (20)$$

Furthermore, we impose non-negativity, orthogonality, smoothness and sparsity constraints on spatial factors to ensure uniqueness. The corresponding M-P diagram is shown in Fig. 14 and new parameters with constraints are found by:

$$(\hat{\mathbf{M}}_C, \hat{\mathbf{M}}_G, \hat{\mathbf{T}}_V, \hat{\mathbf{F}}_V, \hat{\mathbf{M}}_B, \hat{\mathbf{T}}_B)$$
$$= \underset{\mathbf{M}_C, \mathbf{M}_G, \mathbf{T}_V, \mathbf{F}_V, \mathbf{M}_B, \mathbf{T}_B}{\arg\min} \begin{cases} \frac{1}{2} \| \mathcal{S}_T - [\![ \mathbf{K}[\mathbf{M}_C, \mathbf{M}_G], \mathbf{T}_V, \mathbf{F}_V ]\!] \|_2^2 \\ + \gamma \frac{1}{2} \| \mathbf{B} - [\![ [\mathbf{M}_C, \mathbf{M}_B], \mathbf{T}_B ]\!] \|_2^2 \\ + \lambda_1 \| \mathbf{M}_C \|_1 + \frac{1}{2} \lambda_2 \| \mathbf{L} \mathbf{M}_C \|^2 \\ + \lambda_3 \| \mathbf{M}_G \|_1 + \frac{1}{2} \lambda_4 \| \mathbf{L} \mathbf{M}_G \|^2 \\ + \lambda_5 \| \mathbf{M}_B \|_1 + \frac{1}{2} \lambda_6 \| \mathbf{L} \mathbf{M}_B \|^2 \end{cases}$$

s.t. $[\mathbf{M}_C, \mathbf{M}_G]^T [\mathbf{M}_C, \mathbf{M}_G] = \mathbf{I}, [\mathbf{M}_C, \mathbf{M}_B]^T [\mathbf{M}_C, \mathbf{M}_B] = \mathbf{I}$,
$\mathbf{M}_C \geq 0, \ \mathbf{M}_G \geq 0, \mathbf{M}_B \geq 0, \mathbf{F}_V \geq 0$
(21)

The model in (21) can also be interpreted as the estimation of neuronal activity through two sources of information with multiple priors. The $\gamma$ parameter takes into account the scale difference between EEG and fMRI. The M-P Diagram for this type of analysis is provided in Fig. 14.

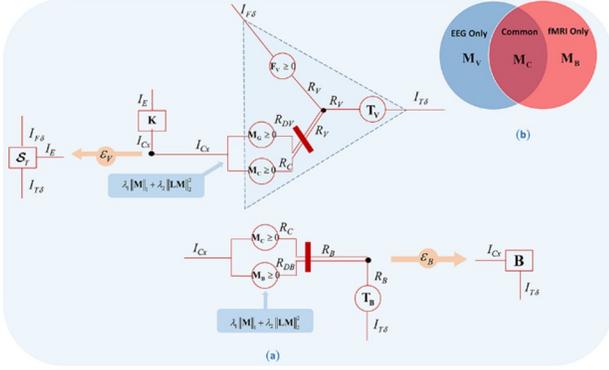

**Fig. 14.** Coupled Matrix-Tensor Factorization. (a) M-P diagram for the coupled matrix-tensor factorization for EEG/fMRI fusion. The EEG tensor $\mathcal{S}_T$ and the fMRI matrix $\mathbf{B}$ are decomposed simultaneously on common and discriminant spatial subspaces to encompass different physiological sources. The spatial signature $\mathbf{M}$ involves common component $\mathbf{M_C}$ and two uncommon $\mathbf{M_G}, \mathbf{M_B}$ components. The fMRI spatial signature is $[\mathbf{M_C}, \mathbf{M_B}]$ and the temporal signature is $\mathbf{T_B}$. For the EEG, the spatial signature of the generators is $[\mathbf{M_C}, \mathbf{M_V}]$, the temporal signature is $\mathbf{T_V}$ and the spectral signature is $\mathbf{F_V}$. By incorporating the lead field matrix $\mathbf{K}$, the model extends the decomposition of EEG to source space. M-P diagrams of EEG and fMRI are separated for a better of visualization. (b) Explicit representation for common and discriminative subspaces are shown. Note that the common subspace is represented with $\mathbf{M_C}$.

We applied the proposed algorithm on a simultaneously recorded EEG - fMRI data [103]. In this experiment, flashing light stimuli in thirteen frequencies in the range of 6 Hz to 42 Hz were presented in a block design paradigm. For this analysis, data segments in the resting periods of the 6 Hz stimulation session of one subject is used. Further information about the acquisition and preprocessing of the data can be found in [103].

The EEG was filtered with a high-pass filter with a cutoff frequency at 60 Hz and segmented in 2981 ms duration segments to match repetition time of fMRI (TR = 2981 ms). Thomson multi-taper method is used to calculate the power spectrum of each segment [104]. We extracted the resting periods of the whole experiment and used for further analysis. This resulted in an EEG tensor $\mathcal{S}_T \in \mathbb{R}^{I_E \times I_S \times I_{F\delta}}$ with $I_E = 31$ channels, $I_{T\delta} = 38$ time points and $I_{F\delta} = 58$ frequency points. The lead field was computed using a realistic head model with three homogenous isotropic conductor boundaries based on the MNI brain atlas.

The fMRI data was normalized to standard MNI space. The voxels on the cortical grid of EEG source space were extracted. Grand mean scaling over the session for the voxels inside the mesh was performed and BOLD values were normalized to obtain a percentage change. At the end, we had an fMRI data matrix $\mathbf{B} \in \mathbb{R}^{I_{Cx} \times I_{T\delta}}$ with $I_{Cx} = 5124$ cortical grid points and $I_{T\delta} = 38$ time points. Being one for common, one for individual EEG and one for individual fMRI atom, in total 3 atoms are extracted from constrained CMTF model.

Fig. 15(I) shows the spatial, temporal and spectral signatures of the common atom. Since the two datasets were coupled only in spatial dimension, two temporal signatures for each modality were obtained—but these show only irregular activity and will not be described further.

The common spatial signature, $\mathbf{M_C}$ shows a clear activation in occipital areas with an EEG spectral signature peaking at 10 Hz—which identifies it with the $\alpha$ activity found by other techniques and shown in Figs. 6, 12 and 13, in line [45]. Pearson's correlation coefficient between the temporal signatures of EEG and fMRI of this common component is found to be -0.3346 with a p-value of 0.04.

The discriminant fMRI atom is shown in Fig. 15(II). Spatial signature showed activation mostly in inferior frontal areas of left and right hemispheres, inferior parietal and middle temporal areas of right hemisphere, precuneus and caudate. When the model order of the fMRI is increased these regions are distributed on separate atoms (results not shown). It seems therefore that the discriminant atom of the fMRI might be the result of the interaction of several of the reported resting state networks.

The discriminant EEG atom shows a *1/f* decay in spectral signature with diffused activations in inferior and middle frontal areas, as well as the temporal areas of both hemispheres which identifies it with the $\xi$ EEG process mentioned in the previous section (see Fig. 15 (III-a,b)).

It seems interesting that the two last techniques, one applied to EEG/DTI data and the other to EEG/fMRI data seem to support the $\xi - \alpha$ model proposed in[99].

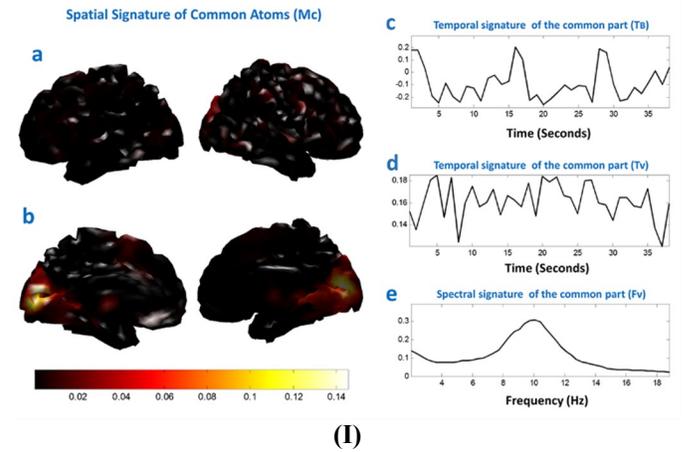

(I)

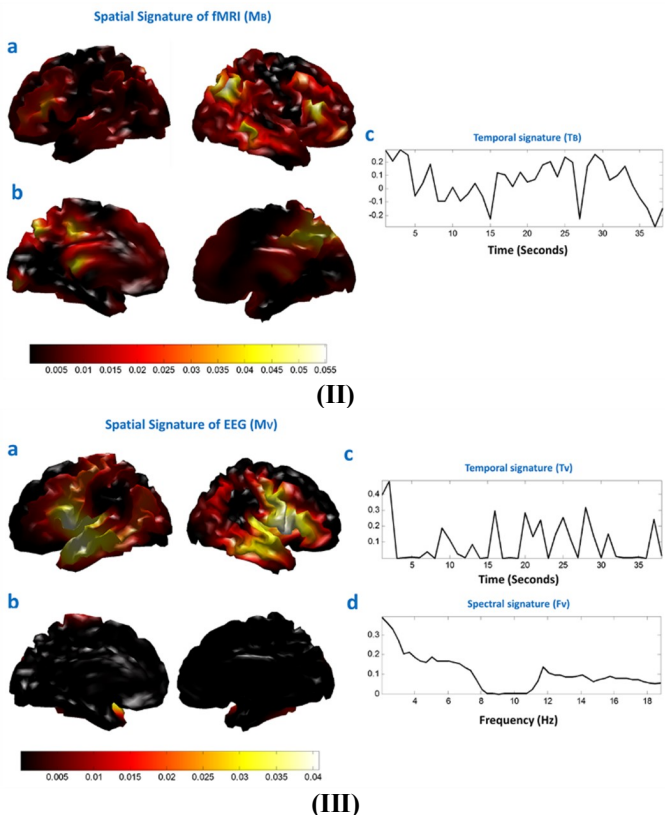

**Fig. 15.** Coupled Matrix-Tensor Factorization on EEG-fMRI data. (I) Common atom extracted from CMTF. (a-b) Spatial signature $(\mathbf{M}_C)$ shows the distribution of activation of the spatial signature on the lateral and medial views of left and right hemispheres. Activity is localized in occipital cortex. (c) fMRI temporal signature of the common atom $(\mathbf{T}_B(:,1))$ (d) EEG temporal signature of the common atom $(\mathbf{T}_V(:,1))$ (e) EEG spectral signature of common atom $(\mathbf{F}_V(:,1))$. 10 Hz peak in the EEG spectral signature indicates an alpha band activity. (II) Discriminant fMRI atom. (a-b) Spatial signature of the discriminant fMRI atom $(\mathbf{M}_B(:,2))$ projected on the lateral and medial views of the left and right hemispheres. (c) Temporal course of the discriminant fMRI atom $(\mathbf{T}_B(:,2))$. fMRI activity is diffuse mostly on the frontal and temporal regions. (III) Discriminant EEG atom. (a-b) Spatial signature of the discriminant EEG atom projected on the lateral and medial views of the left and right hemispheres $(\mathbf{M}_G(:,2))$. A diffused activity is revealed. (c) Temporal signature $(\mathbf{T}_V(:,2))$ (d) Spectral signature $(\mathbf{F}_V(:,2))$. Energy of the spectral signature decreases towards higher frequencies showing the ξ process. Spatial distribution is diffused over temporal and inferior frontal areas. All of the signatures are normalized to unit norm.

### C. Other fields of application of Tensor fusion

By integrating different measurements of a phenomenon we can overcome the lack of precision these offer, providing a complementary vision of things we cannot measure directly.

Many examples of multimodal data integration of different fields are available, such as dynamic clustering by the combination of text and image information on the web [1], speech recognition using complementary hand gesture [105], multisensory image fusion, that is, merging relevant information proceeding from many images, that comes from multiple sensors [106] and human emotion recognition by exploring data arising from speech, face image and thermal image [107]. Multimodal fusion for biological data can be found in [3] for integration of gene expression data with text information and for metabolomics in [108].

## VIII. ALGORITHMS AND SOFTWARE FOR TENSOR BASED PROBLEMS

Throughout this review has concentrated on the graphical properties of tensor diagrams and on the modeling advantages of tensor models. In this section we focus on a more practical subject, that is, the methods and algorithms used for fitting such models. We explain the choices we have made motivated by the main numerical difficulties. We then describe major software packages that are available to implement tensor models.

### A. Optimization of the model functional

Though there are many techniques for tensor network decomposition, we present only the PARAFAC decomposition. Besides ease of interpretation, a major reason for this selection is the ease it may be augmented with multiple penalizers that enhance interpretability and numerical stability. In fact, the alternating least squares algorithm (ALS), often used to estimate PARAFAC model, is an easily implementable and flexible algorithm. ALS is an iterative algorithm in which, at each step only one signature is estimated, considering all others to be fixed. Since each step is a linear regression, penalization methods can be incorporated naturally. ALS has been improved with line search at each step [109], though it can converge slowly, especially when the components are collinear.

Other methods such as gradient based optimization methods [110] and generalized Schur decomposition [111] have been developed as an alternative to overcome the limitations of ALS. In addition, probabilistic methods for general tensor factorizations are presented in [112], [113]. These are alternatives to the algorithmic choices we have made.

Our choice in this article has been a modification of ALS, the Hierarchical Alternating Least Squares algorithm (HALS) in which, at each step of ALS, only one of the component of a factor is estimated, fixing other signatures of all atoms [114]. In particular we have implemented a gradient based HALS for the estimation of parameters for which orthogonality and nonnegativity is required: orthogonality can easily be imposed column wise (refer to [115] for details).

In addition since we use a combination of penalties with different forms of regression and decomposition, we decompose our optimization problem into its additive components and combine the solutions by means of the alternating direction method of multipliers (ADMM) [116][117][118].

The distributed nature of ADMM is also especially geared to solve very large problems in big data since multiple penalization problems may merged into a single global optimization which is solved piece-wise. An example of this this splitting behavior of ADMM is illustrated by (14). Influence fields may be required to be smooth both in incoming (receiver) and outgoing (emitter) directions with a different profile. An analytical solution with these two priors is not possible due to the penalization on two of the dimensions of the connectivity tensor. ADMM shows its strength by splitting this optimization into two tractable sub-problems. Similar approach can be embraced in (17) and (21) for assigning multiple priors on components of a PARAFAC decomposition.

### B. Selection of model parameters

Multiple penalization methods entail tuning of a large amount of hyper-parameters, with scant knowledge about the actual effect of parameters changes.

We have adopted the approach to carry out parameter selection based on the Bayesian Information Criterion (BIC) This requires the estimation of the degrees of freedom (dof), which reflects model complexity, for a given set of the parameters. While general formulations of dof exist [119] and explicit formulas are available for many of the most common regularization problems [120], [121], the complicated nature (e.g. non-differentiability) of some regularizers makes it very hard to find an accurate mathematical expression. For this means, general purpose non-parametric methods have been developed, like [122]. Dof for tensor problems are calculated according to the penalization function applied on each signature e.g. for a smooth lasso type of penalization, the degrees of freedom formulated by [120] is used.

The hyper-parameters chosen are those that minimize BIC, a search over a suitable grid being carried out. We illustrate this procedure by one of the most extreme cases in terms of number of hyper-parameters. This is the CMTF objective function in (21) with 10 hyper-parameters: 6 for smooth lasso type penalization, 3 for the orthogonality constraints on spatial signatures and 1 for the variance difference between two modalities. Hyper-parameters for orthogonality constraints are determined as described in [115]. We present the BIC formulations for this model in Appendix I.

### C. Selection of the number of atoms

PARAFAC type decompositions requires the determination of the number of components or atoms which is also known as the tensor rank. There is not any straightforward algorithm to determine the rank of a specific given tensor; in fact, this problem is NP-hard [123]. Nevertheless practical methods and heuristics have been developed in order to automatically determine the number of components in PARAFAC model such as Corcondia by [124] and [125]. For the determination of model order of PARAFAC based models defined in this article, we used Corcondia and evaluated the explained variance for different number of components. For the selection of the number of components ($R_C$) of CMTF model, a heuristic approach is applied by decomposing two datasets independently. After deciding for an initial value of $R_C$ based on spatial signatures, several model orders are tried and the best model is selected.

### D. Software

There are available software sources for tensor decompositions and tensor operations. The MATLAB Tensor Toolbox offers classes for tensor operations and have several tensor decomposition algorithms [126]. N-way Toolbox supports constrained decompositions and N-PLS [127]. CuBatch is a MATLAB based software for both tensor and classical data analysis, and validation tools providing graphical outputs [128]. Tensorlab provides several decomposition algorithms including structured data fusion of tensors and complex optimization tools [129]. TDALAB and Tensorbox have a graphical user interface supporting several tensor decomposition types [130], [131]. The MATLAB CMTF Toolbox provides constrained and unconstrained algorithms for the joint decomposition of tensors with different orders by using PARAFAC [132]. ERPWAVELAB is a MATLAB based toolbox for multi-channel time-frequency analysis of EEG and MEG by various tensor decompositions including PARAFAC, shifted PARAFAC and Tucker [133]. TT Toolbox includes functions for basic operations with tensor train tensors that are the low-parametric representations of high-dimensional tensors [134]. Python implementation of TT-Toolbox is also available [135]. Tensor Toolbox in Python includes the decomposition of tensors in tensor-train format and spectral tensor-train format [136]. For the construction and manipulation of tensors in the hierarchical tucker format, Hierarchical tucker toolbox is available [137]. TensorReg Toolbox for MATLAB provides sparse PARAFAC and Tucker regression functions [138].

Many of the algorithms developed by the authors can be found at: http://www.cneuro.cu/software/tensor and http://neurosignal.boun.edu.tr/software/tensor.

## IX. CONCLUSIONS

In this paper, we have presented a general framework for tensor analysis of single modality model inversion and multimodal data fusion. We introduce the Markov-Penrose (M-P) diagrams to unify graphical tensor notations with that for Directed Acyclic Graphical (DAG) description of Bayesian statistical models. Using these diagrams different approaches for the solution of inverse problems of EEG and fMRI are described as well as models for their fusion in common domains. Additionally we proposed a tensor MAR for modeling the causal brain networks and reviewed symmetrical fusion methods with the proposed notation. We reviewed algorithms and software packages for implementation of tensor based problems.

EEG and fMRI are mediated by different physiological processes from neural activation leading to differences in their spatial and temporal resolutions. Biophysical models have been addressed to fuse electrical and metabolic signals in meso-scale. Due to the indirect nature of these signals, inverse problems for each modality should be solved to cover the interactions between modalities which are intrinsically ill-posed in their nature. The examples shown with simulations and real data support the usefulness of this type of approach.

As the amount of neuroimaging data increases tremendously, methods dealing with this problem should be developed. Statistical methods based on tensors embraces the high dimensionality of the multimodal data. The M-P tensor notation based on DAGs and Penrose diagrams which unify mathematical models for connectivity and multimodal fusion, may play a heuristic role in suggesting new ways to analyze data, not only in neuroscience, but possibly in other fields.

APPENDIX I

BIC formulations for coupled and uncoupled components of the spatial signatures are given as:

$$\text{BIC}(\mathbf{M_C}) = \frac{1}{2}\|\mathcal{S}_T - [\![\mathbf{K}[\mathbf{M_C}, \mathbf{M_G}], \mathbf{T_V}, \mathbf{F_V}]\!]\|_2^2$$
$$+ \gamma \frac{1}{2}\|\mathbf{B} - [\![[\mathbf{M_C}, \mathbf{M_B}], \mathbf{T_B}]\!]\|_2^2$$
$$+ df(\mathbf{M_C}) \cdot \log(n_V + n_B)/(n_V + n_B)$$

$$\text{BIC}(\mathbf{M_G}) = \frac{1}{2}\|\mathcal{S}_T - [\![\mathbf{K}[\mathbf{M_C}, \mathbf{M_G}], \mathbf{T_V}, \mathbf{F_V}]\!]\|_2^2$$
$$+ df(\mathbf{M_V}) \cdot \log(n_V)/n_V$$

$$\text{BIC}(\mathbf{M_B}) = \frac{1}{2}\|\mathbf{B} - [\![[\mathbf{M_C}, \mathbf{M_B}], \mathbf{T_B}]\!]\|_2^2$$
$$+ df(\mathbf{M_B}) \cdot \log(n_B)/n_B$$

$n_V$ and $n_B$ are the number of elements in $\mathcal{S}_T$ and $\mathbf{B}$, respectively and $df$ is the degrees of freedom computed as in [120]. Hyper-parameters $\lambda_1$ to $\lambda_6$ and $\gamma$ in (21) are found as the minimum of the BIC multidimensional arrays given above.

ACKNOWLEDGMENTS

We wish to thank Fa-Hsuan Lin for providing the fMRI data used in the section on connectivity, Tamer Demiralp for providing the EEG/fMRI data used in the section on CMTF, Jorge Bosch-Bayard for his valuable help in using the STONNICA procedure and Leticia Avila González for help in preparing the figures.

REFERENCES

[1] W. Lu, L. Li, T. Li, H. Zhang, and J. Guo, "Web multimedia object clustering via information fusion," in *Proceedings of the International Conference on Document Analysis and Recognition, ICDAR*, 2011, no. 1, pp. 319–323.

[2] F. Janoos, H. Denli, and N. Subrahmanya, "Multi-scale Graphical Models for Spatio-Temporal Processes," in *Advances in Neural Information Processing Systems 27*, Z. Ghahramani, M. Welling, C. Cortes, N. D. Lawrence, and K. Q. Weinberger, Eds. Curran Associates, Inc., 2014, pp. 316–324.

[3] E. Zeng, C. Yang, T. Li, and G. Narasimhan, "Clustering Genes using Heterogeneous Data Sources," *Int. J. Knowl. Discov. Bioinforma.*, vol. 1, no. 2, pp. 12–28, 2010.

[4] BRAIN Initiative, "BRAIN Initiative," 2015. [Online]. Available: http://braininitiative.nih.gov/.

[5] Human Brain Project, "Human Brain Project," 2015. [Online]. Available: https://www.humanbrainproject.eu/.

[6] T. Jiang, "Brainnetome: A new -ome to understand the brain and its disorders," *Neuroimage*, vol. 80, pp. 263–272, 2013.

[7] H. A. Whiteford, L. Degenhardt, J. Rehm, A. J. Baxter, A. J. Ferrari, H. E. Erskine, F. J. Charlson, R. E. Norman, A. D. Flaxman, N. Johns, R. Burstein, C. J. L. Murray, and T. Vos, "Global burden of disease attributable to mental and substance use disorders: Findings from the Global Burden of Disease Study 2010," *Lancet*, vol. 382, no. 9904, pp. 1575–1586, 2013.

[8] K. Amunts, C. Lepage, L. Borgeat, H. Mohlberg, T. Dickscheid, M.-É. Rousseau, S. Bludau, P.-L. Bazin, L. B. Lewis, A.-M. Oros-Peusquens, N. J. Shah, T. Lippert, K. Zilles, and A. C. Evans, "BigBrain: an ultrahigh-resolution 3D human brain model.," *Science (80-. ).*, vol. 340, no. 6139, pp. 1472–5, 2013.

[9] S. N. Sotiropoulos, S. Jbabdi, J. Xu, J. L. R. Andersson, S. Moeller, E. J. Auerbach, M. F. Glasser, M. Hernandez, G. Sapiro, M. Jenkinson, D. a. Feinberg, E. Yacoub, C. Lenglet, D. C. Van Essen, K. Ugurbil, and T. E. J. Behrens, "Advances in diffusion MRI acquisition and processing in the Human Connectome Project," *Neuroimage*, vol. 80, pp. 125–143, 2013.

[10] K. Uğurbil, J. Xu, E. J. Auerbach, S. Moeller, A. T. Vu, J. M. Duarte-Carvajalino, C. Lenglet, X. Wu, S. Schmitter, P. F. Van de Moortele, J. Strupp, G. Sapiro, F. De Martino, D. Wang, N. Harel, M. Garwood, L. Chen, D. a. Feinberg, S. M. Smith, K. L. Miller, S. N. Sotiropoulos, S. Jbabdi, J. L. R. Andersson, T. E. J. Behrens, M. F. Glasser, D. C. Van Essen, and E. Yacoub, "Pushing spatial and temporal resolution for functional and diffusion MRI in the Human Connectome Project," *Neuroimage*, vol. 80, pp. 80–104, 2013.

[11] N. K. Logothetis, "What we can do and what we cannot do with fMRI.," *Nature*, vol. 453, no. 7197, pp. 869–78, Jun. 2008.

[12] C. M. Michel, M. M. Murray, G. Lantz, S. Gonzalez, L. Spinelli, and R. Grave De Peralta, "EEG source imaging," *Clinical Neurophysiology*, vol. 115, no. 10. pp. 2195–2222, 2004.

[13] D. A. Boas, C. E. Elwell, M. Ferrari, and G. Taga, "Twenty years of functional near-infrared spectroscopy: Introduction for the special issue," *Neuroimage*, vol. 85, pp. 1–5, 2014.

[14] A. W. Toga and J. C. Mazziotta, Eds., *Brain Mapping: The Methods*. Academic Press, 2002.

[15] O. Sporns, *Networks of the Brain*. MIT press, 2011.

[16] P. L. Nunez and R. Srinivasan, *Electric Fields of the Brain: The Neurophysics of EEG*, 2nd ed. Oxford University Press, 2006.

[17] G. Buzsaki, *Rhythms of the Brain*. Oxford University Press, 2006.

[18] K. J. Friston, A. Mechelli, R. Turner, and C. J. Price, "Nonlinear responses in fMRI: the Balloon model, Volterra kernels, and other hemodynamics.," *Neuroimage*, vol. 12, no. 4, pp. 466–477, Oct. 2000.

[19] K. J. Friston, A. M. Bastos, A. Oswal, B. van Wijk, C. Richter, and V. Litvak, "Granger causality revisited," *NeuroImage*, vol. 101, The Authors, pp. 796–808, 05-Jul-2014.

[20] S. Makeig, S. Debener, J. Onton, and A. Delorme, "Mining event-related brain dynamics," *Trends Cogn. Sci.*, vol. 8, no. 5, pp. 204–210, 2004.

[21] S. Makeig and J. Onton, "ERP Features and EEG Dynamics," in *The Oxford Handbook of Event-Related Potential Components*, no. July, E. S. Kappenman and S. J. Luck, Eds. Oxford University Press, 2011.

[22] V. D. Calhoun and T. Adalı, "Multisubject independent component analysis of fMRI: a decade of intrinsic networks, default mode, and neurodiagnostic discovery.," *IEEE Rev. Biomed. Eng.*, vol. 5, pp. 60–73, Jan. 2012.

[23] P. A. Valdés-Sosa, J. M. Sanchez-Bornot, R. C. Sotero, Y. Iturria-Medina, Y. Aleman-Gomez, J. Bosch-Bayard, F. Carbonell, and T. Ozaki, "Model driven EEG/fMRI fusion of brain oscillations.," *Hum. Brain Mapp.*, vol. 30, no. 9, pp. 2701–21, Sep. 2009.

[24] K. Uludağ and A. Roebroeck, "General overview on the merits of multimodal neuroimaging data fusion.," *Neuroimage*, vol. 102, pp. 3–10, May 2014.

[25] J. Sui, T. Adalı, Q. Yu, J. Chen, and V. D. Calhoun, "A review of multivariate methods for multimodal fusion of brain imaging data.," *J. Neurosci. Methods*, vol. 204, no. 1, pp. 68–81, Feb. 2012.

[26] N. M. Correa, T. Eichele, T. Adalı, Y.-O. Li, and V. D. Calhoun, "Multi-set canonical correlation analysis for the fusion of concurrent single trial ERP and functional MRI," *Neuroimage*, vol. 50, no. 4, pp. 1438–45, May 2010.

[27] Y. Levin-Schwartz, V. D. Calhoun, and T. Adalı, "Data-driven fusion of EEG, functional and structural MRI: A comparison of two models," in *48th Annual Conference on Information Sciences and Systems (CISS)*, 2014, pp. 1–6.

[28] J. Sui, H. He, G. D. Pearlson, T. Adalı, K. a Kiehl, Q. Yu, V. P. Clark, E. Castro, T. White, B. a Mueller, B. C. Ho, N. C. Andreasen, and V. D. Calhoun, "Three-way (N-way) fusion of brain imaging data based on mCCA+jICA and its application to discriminating schizophrenia.," *Neuroimage*, vol. 66, pp. 119–32, Feb. 2013.

[29] R. F. Silva, S. Plis, T. Adalı, and V. D. Calhoun, "A statistically motivated framework for simulation of stochastic data fusion models applied to multimodal neuroimaging.," *Neuroimage*, vol. 102, pp. 92–117, Apr. 2014.

[30] X. Lei, P. A. Valdés-Sosa, and D. Yao, "EEG/fMRI fusion based on independent component analysis: integration of data-driven and model-driven methods.," *J. Integr. Neurosci.*, vol. 11, no. 3, pp. 313–37, Sep. 2012.

[31] R. A. Harshman, "Foundations of the PARAFAC procedure: Models and conditions for an 'explanatory' multimodal factor analysis," *UCLA Work. Pap. Phonetics*, vol. 16, pp. 1–84, 1970.

[32] A. Field and D. Graupe, "Topographic component (Parallel Factor) analysis of multichannel evoked potentials: practical issues in trilinear spatiotemporal decomposition," *Brain Topogr.*, vol. 3, no. 4, pp. 407–423, 1991.

[33] J. Möcks, "Topographic components model for event-related potentials and some biophysical considerations.," *IEEE Trans. Biomed. Eng.*, vol. 35, no. 6, pp. 482–4, Jun. 1988.

[34] R. Bro, "Multi-way analysis in the food industry: models, algorithms, and applications," Universiteit van Amsterdam, 1998.

[35] A. Cichocki, R. Zdunek, A. H. Phan, and S. Amari, *Nonnegative Matrix and Tensor Factorizations: applications to exploratory multiway data analysis and blind source separation*. JohnWiley & Sons, Ltd, 2009.

[36] T. G. Kolda and B. W. Bader, "Tensor Decompositions and Applications," *SIAM Rev.*, vol. 51, no. 3, p. 455, 2009.

[37] M. Mørup, "Applications of tensor (multiway array) factorizations and decompositions in data mining," *Wiley Interdiscip. Rev. Data Min. Knowl. Discov.*, vol. 1, no. 1, pp. 24–40, Jan. 2011.

[38] E. Acar and B. Yener, "Unsupervised Multiway Data Analysis: A Literature Survey," *IEEE Trans. Knowl. Data Eng.*, vol. 21, no. 1, pp. 6–20, Jan. 2009.

[39] A. Cichocki, "Tensor Decompositions: A New Concept in Brain Data Analysis?," *arXiv Prepr. arXiv1305.0395*, May 2013.

[40] A. Cichocki, "Era of Big Data Processing: A New Approach via Tensor Networks and Tensor Decompositions," *arXiv Prepr. arXiv1403.2048*, pp. 1–28, Mar. 2014.

[41] A. Cichocki, "Tensor Networks for Big Data Analytics and Large-Scale Optimization Problems," *arXiv Prepr. arXiv1407.3124*, pp. 1–36, 2014.

[42] E. Acar, M. A. Rasmussen, F. Savorani, T. Næs, and R. Bro, "Understanding data fusion within the framework of coupled matrix and tensor factorizations," *Chemom. Intell. Lab. Syst.*, vol. 129, pp. 53–63, Jun. 2013.

[43] F. Miwakeichi, E. Martínez-Montes, P. A. Valdés-Sosa, N. Nishiyama, H. Mizuhara, and Y. Yamaguchi, "Decomposing EEG data into space-time-frequency components using Parallel Factor Analysis.," *Neuroimage*, vol. 22, no. 3, pp. 1035–45, Jul. 2004.

[44] E. Martínez-Montes, J. M. Sánchez-Bornot, and P. A. Valdés-Sosa, "Penalized parafac analysis of spontaneous EEG recordings," *Stat. Sin.*, vol. 18, no. 4, pp. 1149–1464, 2008.

[45] E. Martínez-Montes, P. A. Valdés-Sosa, F. Miwakeichi, R. I. Goldman, and M. S. Cohen, "Concurrent EEG/fMRI analysis by multiway Partial Least Squares.," *Neuroimage*, vol. 22, no. 3, pp. 1023–34, Jul. 2004.

[46] J. M. Sanchez-Bornot, E. Martínez-Montes, A. Lage-Castellanos, M. Vega-Hernández, and P. A. Valdés-Sosa, "Uncovering sparse brain effective connectivity: a voxel-based approach using penalized regression.," *Stat. Sin.*, vol. 18, pp. 1501–1518, 2008.

[47] P. A. Valdés-Sosa, J. M. Sánchez-Bornot, M. Vega-Hernández, L. Melie-García, A. Lage-Castellanos, E. Canales-Rodríguez, M. Valdes-Sosa, and J. M. Bornot-Sánchez, "Granger causality on spatial manifolds: applications to neuroimaging," in *Handbook of Time Series Analysis: Recent Theoretical Developments and Applications*, Wiley-VCH, 2006, pp. 1–53.

[48] P. A. Valdés-Sosa, M. Vega-Hernández, J. M. Sánchez-Bornot, E. Martínez-Montes, and M. A. Bobes, "EEG source imaging with spatio-temporal tomographic nonnegative independent component analysis.," *Hum. Brain Mapp.*, vol. 30, no. 6, pp. 1898–1910, Jun. 2009.

[49] M. Vega-Hernández, E. Martínez-Montes, J. M. Sanchez-Bornot, A. Lage-Castellanos, and P. A. Valdés-Sosa, "Penalized least squares methods for solving the EEG inverse problem," *Stat. Sin.*, vol. 18, no. 4, pp. 1535–1551, 2008.

[50] M. E. Kilmer, K. Braman, N. Hao, and R. C. Hoover, "Third-Order Tensors as Operators on Matrices: A Theoretical and Computational Framework with Applications in Imaging," *SIAM J. Matrix Anal. Appl.*, vol. 34, no. 1, pp. 148–172, 2013.

[51] M. E. Kilmer and C. D. Martin, "Factorization strategies for third-order tensors," *Linear Algebra Appl.*, vol. 435, no. 3, pp. 641–658, Aug. 2011.

[52] R. D. Pascual-Marqui, C. M. Michel, and D. Lehmann, "Low resolution electromagnetic tomography: a new method for localizing electrical activity in the brain," *Int. J. Psychophysiol.*, vol. 18, pp. 49–65, 1994.

[53] J. Bosch-Bayard, P. A. Valdés-Sosa, T. Virues-Alba, E. Aubert-Vazquez, E. R. John, T. Harmony, J. Riera-Diaz, and N. J. Trujillo-Barreto, "3D Statistical Parametric Mapping of EEG Source Spectra by Means of Variable Resolution Electromagnetic Tomography (VARETA)," *Clin. EEG Neurosci.*, vol. 32, no. 2, pp. 47–61, Apr. 2001.

[54] M. D. Plumbley, "Algorithms for nonnegative independent component analysis," *IEEE Trans. Neural Networks*, vol. 14, no. 3, pp. 534–543, 2003.

[55] G. H. Glover, "Deconvolution of impulse response in event-related BOLD fMRI.," *Neuroimage*, vol. 9, no. 4, pp. 416–29, Apr. 1999.

[56] N. J. Trujillo-Barreto, E. Martínez-Montes, L. Melie-García, and P. A. Valdés-Sosa, "A symmetrical Bayesian model for fMRI and EEG/MEG neuroimage fusion," *Int. J. Bioelectromagn.*, vol. 3, no. 1, 2001.

[57] J. Daunizeau, C. Grova, G. Marrelec, J. Mattout, S. Jbabdi, M. Pélégrini-Issac, J.-M. Lina, and M. H. Benali, "Symmetrical event-related EEG/fMRI information fusion in a variational Bayesian framework.," *Neuroimage*, vol. 36, no. 1, pp. 69–87, May 2007.

[58] T. Murta, M. Leite, D. W. Carmichael, P. Figueiredo, and L. Lemieux, "Electrophysiological correlates of the BOLD signal for EEG-informed fMRI," *Hum. Brain Mapp.*, vol. 36, no. 1, pp. 391–414, 2015.

[59] V. D. Calhoun, L. Wu, K. a. Kiehl, T. Eichele, and G. D. Pearlson, "Aberrant processing of deviant stimuli in schizophrenia revealed by fusion of fMRI and EEG data," *Acta Neuropsychiatr.*, vol. 22, no. 3, pp. 127–138, 2010.

[60] L. De Lathauwer, B. De Moor, and J. Vandewalle, "A Multilinear Singular Value Decomposition," *SIAM J. Matrix Anal. Appl.*, vol. 21, no. 4, pp. 1253–1278, Jan. 2000.

[61] H. Becker, L. Albera, P. Comon, M. Haardt, G. Birot, F. Wendling, M. Gavaret, C. G. Bénar, and I. Merlet, "EEG extended source localization: tensor-based vs. conventional methods.," *Neuroimage*, vol. 96, pp. 143–57, Aug. 2014.

[62] R. Penrose, "Applications of negative dimensional tensors," in *Combinatorial Mathematics and its Applications*, London: Academic Press, 1971, pp. 221–244.

[63] T. Huckle, K. Waldherr, and T. Schulte-Herbrüggen, "Computations in quantum tensor networks," *Linear Algebra Appl.*, vol. 438, no. 2, pp. 750–781, Jan. 2013.

[64] S. Holtz, T. Rohwedder, and R. Schneider, "The Alternating Linear Scheme For Tensor Optimization in the Tensor Train Format," *SIAM J. Sci. Comput.*, vol. 34, no. 2, pp. 683–713, 2012.

[65] T. G. Kolda, "Multilinear operators for higher-order decompositions," Albuquerque, NM, Livermore, CA, 2006.

[66] J. B. Kruskal, "Three-way arrays: rank and uniqueness of trilinear decompositions, with application to arithmetic complexity and statistics," *Linear Algebra and its Applications*, vol. 18, no. 2. pp. 95–138, 1977.

[67] C. M. Bishop, *Pattern Recognition and Machine Learning*. Springer, 2006.

[68] A. Critch, "Algebraic Geometry of Hidden Markov and Related Models," University of California, Berkeley, 2013.

[69] A. Critch and J. Morton, "Algebraic Geometry of Matrix Product States," *Symmetry, Integr. Geom. Methods Appl.*, vol. 10, Sep. 2014.

[70] F. Cong, Q. Lin, L. Kuang, X. F. Gong, P. Astikainen, and T. Ristaniemi, "Tensor decomposition of EEG signals: A brief review," *J. Neurosci. Methods*, 2015.

[71] X. F. Gong, C. Wang, Y. Hao, and Q. Lin, "Combined independent component analysis and canonical polyadic decomposition via joint diagonalization," in *2014 IEEE China Summit & International Conference on Signal and Information Processing (ChinaSIP)*, 2014, no. 1, pp. 804–808.

[72] M. De Vos, D. Nion, S. Van Huffel, and L. De Lathauwer, "A combination of parallel factor and independent component analysis," *Signal Processing*, vol. 92, no. 12, pp. 2990–2999, 2012.

[73] C. F. Beckmann and S. M. Smith, "Tensorial extensions of independent component analysis for multisubject FMRI analysis.," *Neuroimage*, vol. 25, no. 1, pp. 294–311, Mar. 2005.

[74] X. F. Gong and Q. H. Lin, "Spatially constrained parallel factor analysis for semi-blind beamforming," in *Proceedings - 2011 7th International Conference on Natural Computation, ICNC 2011*, 2011, vol. 1, pp. 416–420.

[75] P. A. Valdés-Sosa, A. Roebroeck, J. Daunizeau, and K. J. Friston, "Effective connectivity: Influence, causality and biophysical modeling.," *Neuroimage*, vol. 58, no. 2, pp. 339–361, Apr. 2011.

[76] J. Pearl, "Statistics and causal inference: A review," *Test*, vol. 12, no. 2. pp. 281–345, 2003.

[77] N. Wiener, "The theory of prediction," in *Modern Mathematics for Engineers*, E. F. Beckenbach, Ed. New York: McGraw-Hill, 1956.

[78] H. Akaike, "On the use of a linear model for the identification of feedback systems," *Ann. Inst. Stat. Math.*, vol. 20, no. 1, pp. 425–439, 1968.

[79] C. W. J. Granger, "Economic processes involving feedback," *Inf. Control*, vol. 6, no. 1, pp. 28–48, 1963.

[80] T. Schweder, "Composable markov processes," *J. Appl. Probab.*, vol. 7, no. 2, pp. 400–410, 1970.

[81] M. Eichler, "A graphical approach for evaluating effective connectivity in neural systems.," *Philos. Trans. R. Soc. Lond. B. Biol. Sci.*, vol. 360, no. 1457, pp. 953–967, 2005.

[82] D. Commenges and A. Gégout-Petit, "A general dynamical statistical model with causal interpretation," *J. R. Stat. Soc. Ser. B Stat. Methodol.*, vol. 71, no. 3, pp. 719–736, 2009.

[83] K. Friston, "Hierarchical models in the brain," *PLoS Comput. Biol.*, vol. 4, no. 11, 2008.

[84] M. Vinck, L. Huurdeman, C. A. Bosman, P. Fries, F. P. Battaglia, C. M. A. Pennartz, and P. H. Tiesinga, "How to detect the Granger-causal flow direction in the presence of additive noise?," *Neuroimage*, vol. 108, pp. 301–318, 2015.

[85] F. H. Lin, J. Ahveninen, T. Raij, T. Witzel, Y. H. Chu, I. P. Jääskeläinen, K. W. K. Tsai, W. J. Kuo, and J. W. Belliveau, "Increasing fMRI sampling rate improves Granger causality estimates," *PLoS One*, vol. 9, no. 6, pp. 26–29, 2014.

[86] A. K. Seth, A. B. Barrett, and L. Barnett, "Granger Causality Analysis in Neuroscience and Neuroimaging," *J. Neurosci.*, vol. 35, no. 8, pp. 3293–3297, 2015.

[87] P. A. Valdés-Sosa, J. M. Sanchez-Bornot, A. Lage-Castellanos, M. Vega-Hernández, J. Bosch-Bayard, L. Melie-García, and E. Canales-Rodríguez, "Estimating brain functional connectivity with sparse multivariate autoregression.," *Philos. Trans. R. Soc. Lond. B. Biol. Sci.*, vol. 360, no. 1457, pp. 969–81, May 2005.

[88] P. Bühlmann and S. van de Geer, *Statistics for High-Dimensional Data*, 1st ed. Springer-Verlag Berlin Heidelberg, 2011.

[89] J. Songsiri, "Graphical Models of Time Series: Parameter Estimation and Topology Selection," University of California, Los Angeles, 2010.

[90] O. Semerci, N. Hao, M. E. Kilmer, and E. L. Miller, "Tensor-based formulation and nuclear norm regularization for multienergy computed tomography," *IEEE Trans. Image Process.*, vol. 23, no. 4, pp. 1678–1693, 2014.

[91] C. R. Dietrich and G. N. Newsam, "Fast and Exact Simulation of Stationary Gaussian Processes through Circulant Embedding of the Covariance Matrix," *SIAM Journal on Scientific Computing*, vol. 18, no. 4. pp. 1088–1107, 1997.

[92] H. Huang, C. Ding, D. Luo, and T. Li, "Simultaneous tensor subspace selection and clustering: The Equivalence of High Order SVD and K-Means Clustering," in *Proceeding of the 14th ACM SIGKDD international conference on Knowledge discovery and data mining - KDD 08*, 2008, p. 327.

[93] C. Ding, T. Li, and M. I. Jordan, "Convex and semi-nonnegative matrix factorizations.," *IEEE Trans. Pattern Anal. Mach. Intell.*, vol. 32, no. 1, pp. 45–55, Jan. 2010.

[94] E. E. Papalexakis, N. D. Sidiropoulos, and R. Bro, "From K-Means to Higher-Way Co-Clustering: Multilinear Decomposition With Sparse Latent Factors," *IEEE Trans. Signal Process.*, vol. 61, no. 2, pp. 493–506, Jan. 2013.

[95] G. R. Murthy, "Concurrent Cyber Physical Systems:Tensor State Space Representation," *arXiv Prepr. arXiv1303.1597*, Mar. 2013.

[96] R. I. Goldman, J. M. Stern, J. Engel, and M. S. Cohen, "Simultaneous EEG and fMRI of the alpha rhythm.," *Neuroreport*, vol. 13, no. 18, pp. 2487–92, Dec. 2002.

[97] D. Mantini, M. G. Perrucci, C. Del Gratta, G. L. Romani, and M. Corbetta, "Electrophysiological signatures of resting state networks in the human brain.," *Proc. Natl. Acad. Sci. U. S. A.*, vol. 104, no. 32, pp. 13170–13175, 2007.

[98] P. A. Valdés-Hernández, A. Ojeda-González, E. Martínez-Montes, A. Lage-Castellanos, T. Virués-Alba, L. Valdés-Urrutia, and P. A. Valdés-Sosa, "White matter architecture rather than cortical surface area correlates with the EEG alpha rhythm.," *Neuroimage*, vol. 49, no. 3, pp. 2328–39, Feb. 2010.

[99] P. A. Valdés-Sosa, J. Bosch-Bayard, R. Grave, J. Hernandez, J. Riera, R. Pascual, R. Biscay, J. Bosch, R. Grave, J. Hernandez, J. Riera, R. Pascual, R. Biscay, J. Bosch-Bayard, R. Grave, J. Hernandez, J. Riera, R. Pascual, and R. Biscay, "Frequency domain models of the EEG," *Brain Topogr.*, vol. 4, no. 4, pp. 309–319, 1992.

[100] R. D. Pascual-Marqui, P. a Valdes-Sosa, and a Alvarez-Amador, "A parametric model for multichannel EEG spectra.," *Int. J. Neurosci.*, vol. 40, no. 1–2, pp. 89–99, 1988.

[101] W. Liu, J. Chan, J. Bailey, C. Leckie, and K. Ramamohanarao, "Mining Labelled Tensors by Discovering both their Common and Discriminative Subspaces," in *Proceedings of the 2013 SIAM Conference on Data Mining*, 2013.

[102] M. J. Rosa, J. Daunizeau, and K. J. Friston, "EEG-fMRI integration: A critical review of biophysical modeling and data analysis approaches," *J. Integr. Neurosci.*, vol. 09, no. 04, pp. 453–476, Dec. 2010.

[103] A. Bayram, Z. Bayraktaroglu, E. Karahan, B. Erdogan, B. Bilgic, M. Ozker, I. Kasikci, A. D. Duru, A. Ademoglu, C. Oztürk, K. Arikan, N. Tarhan, and T. Demiralp, "Simultaneous EEG/fMRI analysis of the resonance phenomena in steady-state visual evoked responses.," *Clin. EEG Neurosci.*, vol. 42, no. 2, pp. 98–106, Apr. 2011.

[104] D. J. Thomson, "Spectrum estimation and harmonic analysis," *Proc. IEEE*, vol. 70, no. 9, pp. 1055–1096, 1982.

[105] M. Miki, N. Kitaoka, C. Miyajima, T. Nishino, and K. Takeda, "Improvement of multimodal gesture and speech recognition performance using time intervals between gestures and accompanying speech," *EURASIP J. Audio, Speech, Music Process.*, vol. 2014, no. 1, p. 2, 2014.

[106] S. Li and B. Yang, "Hybrid multiresolution method for multisensor multimodal image fusion," *IEEE Sens. J.*, vol. 10, no. 9, pp. 1519–1526, 2010.

[107] Y. Yoshitomi, S. I. Kim, T. Kawano, and T. Kitazoe, "Effect of sensor fusion for recognition of emotional states using voice, face image and thermal image of face," in *Proceedings - IEEE International Workshop on Robot and Human Interactive Communication*, 2000, pp. 178–183.

[108] K. Van Deun, T. F. Wilderjans, R. a van den Berg, A. Antoniadis, and I. Van Mechelen, "A flexible framework for sparse simultaneous component based data integration.," *BMC Bioinformatics*, vol. 12, no. 1, p. 448, Jan. 2011.

[109] P. Comon, X. Luciani, and A. L. F. de Almeida, "Tensor decompositions, alternating least squares and other tales," *J. Chemom.*, vol. 23, no. 7–8, pp. 393–405, Jul. 2009.

[110] E. Acar, D. M. Dunlavy, and T. G. Kolda, "A scalable optimization approach for fitting canonical tensor decompositions," *J. Chemom.*, vol. 25, no. 2, pp. 67–86, Feb. 2011.

[111] L. De Lathauwer, B. De Moor, and J. Vandewalle, "Computation of the Canonical Decomposition by Means of a Simultaneous Generalized Schur Decomposition," *SIAM J. Matrix Anal. Appl.*, vol. 26, no. 2, pp. 295–327, Jan. 2004.

[112] Y. K. Yılmaz and A. T. Cemgil, "Algorithms for probabilistic latent tensor factorization," *Signal Processing*, vol. 92, no. 8, pp. 1853–1863, Aug. 2012.

[113] M. Schmidt and S. Mohamed, "Probabilistic Non-negative Tensor Factorisation using Markov Chain Monte Carlo," in *17th European Signal Processing Conference*, 2009.

[114] A. Cichocki, R. Zdunek, and S. Amari, "Hierarchical ALS algorithms for nonnegative matrix and 3D tensor factorization," in *Independent Component Analysis, ICA07*, 2007, pp. 169–176.

[115] K. Kimura, Y. Tanaka, and M. Kudo, "A Fast Hierarchical Alternating Least Squares Algorithm for Orthogonal Nonnegative Matrix Factorization," in *JMLR: Workshop and Conference Proceedings*, 2014, pp. 129–141.

[116] S. Boyd, N. Parikh, E. Chu, B. Peleato, and J. Eckstein, "Distributed Optimization and Statistical Learning via the Alternating Direction Method of Multipliers," *Found. Trends® Mach. Learn.*, vol. 3, no. 1, pp. 1–122, 2010.

[117] A. P. Liavas and N. D. Sidiropoulos, "Parallel Algorithms for Constrained Tensor Factorization via the Alternating Direction Method of Multipliers," *arXiv Prepr. arXiv1409.2383*, pp. 1–26, Sep. 2014.

[118] S. Gandy, B. Recht, and I. Yamada, "Tensor completion and low-n-rank tensor recovery via convex optimization," *Inverse Probl.*, vol. 27, no. 2, p. 025010, Feb. 2011.

[119] C. M. Stein, "Estimation of the Mean of a Multivariate Normal Distribution," *Ann. Stat.*, vol. 9, no. 6, pp. 1135–1151, 1981.

[120] M. Hebiri, "Regularization with the smooth-lasso procedure," *arXiv Prepr. arXiv0803.0668*, 2008.

[121] H. Zou, T. Hastie, and R. J. Tibshirani, "On the 'degrees of freedom' of the lasso," *Ann. Stat.*, vol. 35, no. 5, pp. 2173–2192, Oct. 2007.

[122] J. Ye, "On Measuring and Correcting the Effects of Data Mining and Model Selection," *J. Am. Stat. Assoc.*, vol. 93, no. 441, pp. 120–131, 1998.

[123] J. Håstad, "Tensor rank is NP-complete," *Journal of Algorithms*, vol. 11, no. 4. pp. 644–654, 1990.

[124] R. Bro and H. A. L. Kiers, "A new efficient method for determining the number of components in PARAFAC models," *J. Chemom.*, vol. 17, no. 5, pp. 274–286, Jun. 2003.

[125] M. Mørup and L. K. Hansen, "Automatic relevance determination for multi-way models," *J. Chemom.*, vol. 23, no. 7–8, pp. 352–363, Jul. 2009.

[126] B. W. Bader and T. G. Kolda, "MATLAB Tensor Toolbox." 2015.

[127] C. A. Andersson and R. Bro, "The N-way Toolbox for MATLAB," *Chemom. Intell. Lab. Syst.*, vol. 52, no. 1, pp. 1–4, Aug. 2000.

[128] S. Gourvénec, G. Tomasi, C. Durville, E. Di Crescenzo, C. A. Saby, D. L. Massart, R. Bro, and G. Oppenheim, "CuBatch." 2005.

[129] L. Sorber, M. Van Barel, and L. De Lathauwer, "Tensorlab." 2014.

[130] G. Zhou and A. Cichocki, "TDALAB: Tensor Decomposition Laboratory." 2013.

[131] A. H. Phan, P. Tichavsk, and A. Cichocki, "Tensor-box: a matlab package for tensor decomposition." 2013.

[132] E. Acar, "The MATLAB CMTF Toolbox." 2014.

[133] M. Mørup, "ERPWAVELAB." 2006.

[134] I. V Oseledets, "TT Toolbox." 2009.

[135] I. V Oseledets, T. Saluev, D. V Savostyanov, and S. V Dolgov, "ttpy." 2014.

[136] D. Bigoni, "Tensor Toolbox." 2015.

[137] C. Tobler and D. Kressner, "Hierarchical Tucker Toolbox." 2013.

[138] H. Zhou, "TensorReg Toolbox for Matlab." 2013.

[139] M. Bringas, I. Pedroso, V. Perez, J. Sanchez-Bornot, and P. Valdes-Sosa, "Resting state frequency domain Tomographic ICA," in *XXVII Annual International Meeting for Human Brain Mapping*, 2011.

[140] R. Plonsey and D. B. Heppner, "Considerations of quasi-stationarity in electrophysiological systems," *Bull. Math. Biophys.*, vol. 29, pp. 657–664, 1967.